\documentclass[usenatbib]{mn2e}
\usepackage{graphicx}
\usepackage{picture}

\title[Metallicity Effects on Globular Cluster Sizes]{$N$-body models of globular clusters: metallicity, half-light radii and mass-to-light ratios}
\author[Anna C. Sippel, Jarrod R. Hurley, Juan P. Madrid, William E. Harris]{Anna C. Sippel$^{1}$\thanks{E-mail: asippel@astro.swin.edu.au}, Jarrod R. Hurley$^{1}$, Juan P. Madrid$^{1}$, William E. Harris$^{2}$\\
$^{1}$Centre for Astrophysics and Supercomputing, Swinburne University of Technology, PO Box 218, Hawthorn, VIC 3122, Australia\\
$^{2}$Department of Physics and Astronomy, McMaster University, Hamilton, ON L8S 4M1, Canada}
\begin{document}

\date{accepted for publication in MNRAS}
\maketitle

\begin{abstract}
Size differences of $\approx 20\%$ between red (metal-rich) and blue (metal-poor) sub-populations of globular clusters have been observed, generating an ongoing debate as to weather these originate from projection effects or the difference in metallicity. We present direct $N$-body simulations of metal-rich and metal-poor stellar populations evolved to study the effects of metallicity on cluster evolution. The models start with $N = 100\,000$ stars and include primordial binaries. We also take metallicity dependent stellar evolution and an external tidal field into account. We find no significant difference for the half-mass radii of those models, indicating that the clusters are structurally similar. However, utilizing observational tools to fit half-\emph{light} (or effective) radii confirms that metallicity effects related to stellar evolution combined with dynamical effects such as mass segregation produce an apparent size difference of $17\%$ on average. The metallicity effect on the overall cluster luminosity also leads to higher mass-to-light ratios for metal-rich clusters.
\end{abstract}

\begin{keywords}
globular clusters: general - galaxies: star clusters: general - stars: mass-loss - stars: luminosity function, mass function - methods: $N$-body simulations
\end{keywords}


\section{INTRODUCTION}\label{intro}
\begin{figure}
\centering
\includegraphics[width=0.47\textwidth]{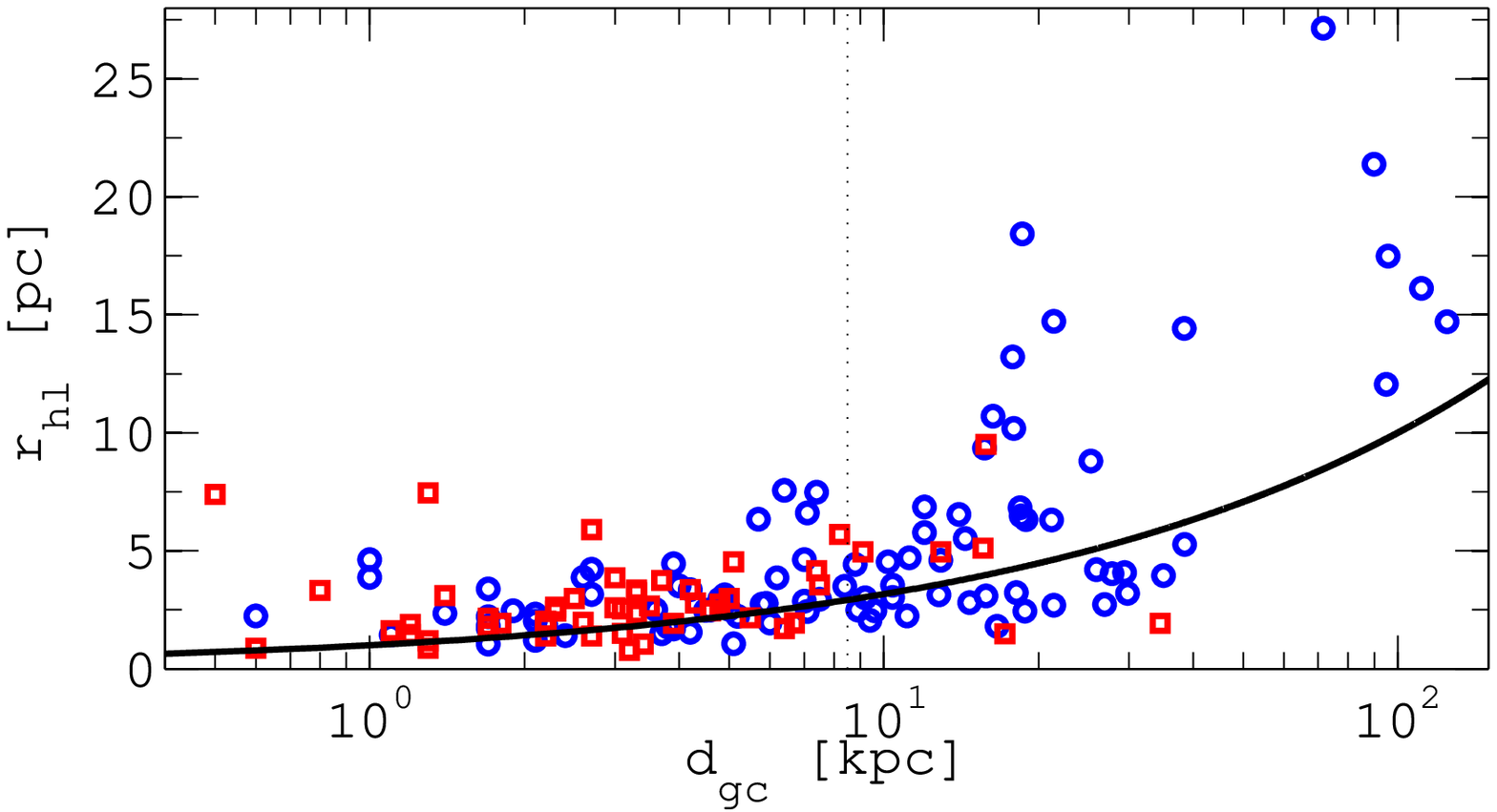}
\caption{Size and galactocentric distance of the MW GC population (compiled from \citealt{Harris2010ed}). Blue circles are used for metal-poor and red squares for metal-rich clusters, the distinction is made at [Fe/H]$=-1.1$. The solid black like denotes the size-distance relation $r_{\rm{hl}} \approx \sqrt{d_{\rm{gc}}}$ from \citet{vdBergh1991}. Metal-poor clusters tend to have larger galactocentric distances as well as larger sizes (half-light radii). The models used for this study are evolved at a galactocentric distance of $8.5\,$kpc, marked by the vertical dotted line.}\label{fig:fig1}
\end{figure}

Globular clusters (GCs) are substantial components of galaxies and found in populations of up to thousands in giant elliptical galaxies \citep{Peng2011}. The Milky-Way (MW) hosts a GC population of $157$ confirmed clusters (\citealt{Harris1996}, 2010 edition), with new clusters still being discovered (e.g. \citealt{Minniti2011}). These clusters live within the bulge as well as the halo of the Galaxy and can - in contrast to star clusters beyond the Local Group - easily be resolved in ground-based observations.
In general, the GC systems of galaxies tend to appear in two sub-populations: a blue and a red component \citep{Zinn1985}. Although the metallicity cannot be inferred directly from the cluster colour due to the age-metallicity degeneracy \citep{Worthey1994}, it has been well accepted that the blue clusters are metal-poor, whereas the red ones are metal-rich. Both sub-populations are old (e.g. \citealt{MarinFranch2009}), with a trend for the red clusters to be more centrally concentrated within their host galaxy's potential than their blue counterparts (\citealt{Kinman1959, BS2006}, also see Fig. \ref{fig:fig1}). 

The ability of the Hubble Space Telescope to partially resolve globular clusters even beyond the Local Group has lead to the finding that i) GCs have mean half-light radii $r_{\rm{hl}}=3\,$pc \citep{Jordan2005} and ii) red clusters are on average $\approx 17-30\%$ smaller than their metal poor counterparts \citep{Kundu1998, Jordan2005, Woodley2010}. Several explanations for this phenomena have been proposed: projection effects and the influence of the tidal field \citep{Larsen2001} or a combined effect of mass segregation and the dependence of main-sequence lifetimes on metallicity \citep{Jordan2004, Jordan2005}. 
Whether either of those effects are dominating or a combination of both can only be investigated through direct star cluster simulations where three-dimensional galactocentric distances are known and stellar evolution is included in the dynamical evolution of the cluster. 

The effects of metallicity on the evolution of a single star manifests itself as a different rate of stellar evolution, which is accompanied by a different mass-loss rate and hence ultimately affects the stars lifetime and remnant mass (see Section \ref{Zeff}). In general, low metallicity stars evolve \emph{faster} along the main sequence than their high metallicity counterparts \citep{SSE}. For a bound system such as a star cluster, the increased mass-loss rate can lead to a lower cluster mass and hence a lower escape velocity and. This in turn can produce a stronger increase in radius for the metal-poor cluster. At later stages, this might also lead to postponed core-collapse for the low metallicity cluster. Both effects could lead to a larger measured cluster size. A preliminary study along these lines has been carried out by \citet{Hurley2004} for open clusters. They showed that an increased escape rate for the metal-rich clusters owing to earlier core-collapse acts to cancel these effects resulting in only a $10 \%$ difference in cluster lifetime for metal-poor versus metal-rich cases - within the statistical noise of fluctuating results from one simulation to another. However, several aspects of our new simulations differ from this preliminary study. Among those are an adjusted binary fraction for GCs and an improved tidal field. Most importantly we also use a higher initial number of stars $N_{\rm{i}}$, bringing the $N$-body models into the GC regime. This ultimately leads to an increase in cluster lifetime and hence not necessarily core-collapse or depletion of stars within a Hubble time. 

In this work, we make use of a set of star cluster simulations evolved with the direct $N$-body code \texttt{NBODY6} \citep{Aarseth1999, Aarseth2003} to study the effects of metallicity on star cluster dynamics, evolution and size (i.e. effective radius) to answer the question if metallicity alone could reproduce the observed size difference. We measure the sizes of these clusters along their evolutionary track with methods used both in observations and theory.

Recently \citet{Downing2012} has published a set of Monte-Carlo models exploring the origin of the observed size difference between metal-rich and metal-poor GCs, which provides an excellent comparison for our work. This follows on from the $N$-body models of \citet{Schulman2012}, who investigated the evolution of half-mass radius with metallicity in small-$N$ clusters. Similarly to \citet{Downing2012}, we shall be careful to make a distinction between the actual size of a star cluster, represented by the half-mass radius (which we shall denote as $r_{\rm{50\%}}$, i.e. the $50\%$ Lagrangian radius), and the observationally determined size (the half-light or effective radius, $r_{\rm{eff}}$).

This paper is structured as follows. We introduce the differences in stellar evolution depending on metallicity in the next Section. In Section \ref{method}, we describe our simulation method and the models we have chosen to evolve. In Section \ref{evolution}, we analyze the evolution of cluster mass, binary fraction, luminosity, half-light radius and mass-to-light ratio which is followed by discussion and conclusions. 
%
%
\section{METALLICITY EFFECTS ON STELLAR AND STAR CLUSTER EVOLUTION}\label{Zeff}
The main sequence (MS) lifetime of a single star depends mainly on its mass (and hence luminosity), but also on its chemical composition: the metallicity $Z$ (or [Fe/H]). \citet{Clayton1968} shows that the MS lifetime can be represented as:
\begin{equation}\label{MSlifetime1}
t_{\rm{MS}} \propto \frac{1}{X} \frac{m/m_{\odot}}{L/L_{\odot}}\,,
\end{equation}
where $m$ and $L$ are a star's mass and luminosity and $X$ the hydrogen fraction. A star's mass at given luminosity scales as:
\begin{equation}\label{MS_M_L}
m \propto \frac{\kappa_0 ^{0.2}}{\mu^{1.4}}\, ,
\end{equation}
where $\kappa_0$ is the central opacity and $\mu$ the mean molecular weight. The hydrogen fraction $X$ and helium abundance $Y$ can be set as a function of metallicity according to: 
\begin{equation}\label{hyd_frac}
X=0.76-3Z
\end{equation}
\begin{equation}\label{hel_frac}
Y=0.24+2Z
\end{equation}
as in \citet{Pols1998}. If $Z$ is decreased, $X$ increases while $Y$ decreases slightly, leading to a marginally lower mean molecular weight: 
\begin{equation}\label{mu}
\mu \approx \frac{2}{1+3X+0.5Y} \,.
\end{equation} 
To first order, it can be assumed that the central opacity is proportional to Z: $\kappa_0 \propto Z$ \citep{Clayton1968}. Using this, in combination with Eqs. \ref{MSlifetime1} and \ref{MS_M_L} we find:
\begin{equation}\label{MSlifetime2}
t_{\rm{MS}} \propto \frac{\kappa_0^{0.2} X}{\mu^{1.4}} \approx \frac{Z^{0.2} X}{\mu^{1.4}}\,,
\end{equation}
with $\kappa_0^{0.2} \propto Z^{0.2}$ being the dominant term in this equation. A lower opacity implies less resistance for escaping photons from the hydrogen burning core and hence a higher luminosity and therefore a shorter lifetime (see also Table \ref{tb:tb1}). For an extended discussion we refer to \citet{Clayton1968}.

\begin{table*}
\centering
 \caption{Main sequence lifetimes for stars with different metallicities. Metallicity $Z$ and [Fe/H] are in the first two columns, followed by the hydrogen ($X$) and helium ($Y$) mass fraction (Eq. \ref{hyd_frac} and \ref{hel_frac}). Even though the mean molecular weight $\mu$ (Eq. \ref{mu}) in column $5$ is barely affected by the metallicity, different relative MS lifetimes $t_{\rm{MS}}$ (column $6$) are caused by a change in opacity for different metallicities according to Eq. \ref{MSlifetime2} . The expected MS lifetimes up to the Hertzsprung Gap according to \citet{SSE} for stars with initially $3$, $1.5$ and $0.8\,M_{\odot}$ are given in the next three columns, followed by the MS turnover mass $m_{\rm{TO}}$ in columns $10$ and $11$ at ages of $11$ and $12\,$Gyr. We note that stars with $Z=0.001$ and $Z=0.0001$ evolve in a similar fashion compared to the metal rich case - hence $Z=0.001$ is also a metal-poor case. This has already been noted by \citet{Hurley2004}, as well as the fact that stars and clusters with $Z=0.01$ evolve similar to solar metallicity $Z=0.02$.}\label{tb:tb1}
  \begin{tabular}{@{}rcccccccc|ccc@{}}
     $Z$       & [Fe/H]  &   $X$    &  $Y$     & $\mu$ & $t_{\rm{MS}}$ (Eq. \ref{MSlifetime2}) & $t_{\rm{MS}}(3\,M_{\odot})$ & $t_{\rm{MS}}(1.5\,M_{\odot})$ & $t_{\rm{MS}}(0.8\,M_{\odot})$ & $m_{\rm{TO}}$ ($11\,$Gyr) & $m_{\rm{TO}}$ ($12\,$Gyr)\\ \hline
  $0.0001$ & $-2.3$ & $0.76$ & $0.24$ & $0.59$ &  $0.25$ & $0.29\,$Gyr & $1.6\,$Gyr & $13.5\,$Gyr  & $0.84\,M_{\odot}$ & $0.83\,M_{\odot}$\\
    $0.001$ & $-1.3$ & $0.76$ & $0.25$ & $0.59$ &  $0.40$ & $0.29\,$Gyr & $1.7\,$Gyr & $14\,$Gyr  & $0.85\,M_{\odot}$ & $0.83\,M_{\odot}$\\
      $0.01$ & $-0.3$ & $0.74$ & $0.26$ & $0.60$ &  $0.60$ & $0.35\,$Gyr & $2.4\,$Gyr & $21.7\,$Gyr  & $0.95\,M_{\odot}$ & $0.91\,M_{\odot}$\\
 \end{tabular}
 \end{table*}

In the $N$-body models, we evolve stars according to the stellar evolution prescriptions of \citet{SSE}, which are based on the detailed models of \citet{Pols1998}. These prescriptions are accurate for a wide range of metallicities and cover all phases of stellar evolution. This means stars are evolved from the zero-age main sequence up to and including the remnant phases: white dwarfs (WDs), neutron stars (NSs) and black holes (BHs). If necessary, the stellar evolutionary track evolves via the giant branch, core helium burning and thermally pulsating asymptotic giant branch. As shown by \citet{SSE}, the difference in MS lifetime is most prominent for low-mass stars and steadily decreases towards higher mass stars until $M \approx 8\,M_{\odot}$, where the high metallicity stars begin to have a shorter MS lifetime, although only marginally (and noting that model uncertainties are more prevalent at higher masses). This implies, that for clusters of the same age, the mass of the most massive MS star (and hence MS turnoff mass $m_{\rm{TO}}$) is higher in a high-$Z$ cluster. Examples for $m_{\rm{TO}}$ are given in Table \ref{tb:tb1}. It is not only the MS lifetime that is altered by the metallicity, but also the remnant mass. For initial masses less than $50\,M_{\odot}$ our models give a maximum black hole mass $m_{\rm{BH}} \approx 28\,M_{\odot}$ for metal-poor stars versus $m_{\rm{BH}} \approx 12\,M_{\odot}$ for metal rich progenitors \citep{Belczynski2006}. This trend is the same for all remnants: a $2\,M_{\odot}$ progenitor with $Z=0.0001$ will end life as a WD of mass $m=0.84\,M_{\odot}$, while a metal-rich counterpart with $Z=0.01$ will have a WD mass of $m=0.66\,M_{\odot}$. This occurs after $\approx 0.9$ and $1.4$ Gyr, respectively. Hence the remnant mass in a metal-poor cluster is always expected to be higher (see also Table \ref{tb:tb2}).

Since there is no strong evidence for an explicit metallicity dependence of the mass-loss rate of giants \citep{Iben1983, Carraro1996, Schroeder2005}, generally mass-loss from the envelope during the giant branch phase and beyond is implemented according to Reimer's law (formula of \citealt{Kudritzki1978}):
\begin{equation}
\dot{m} \propto \frac{L R}{m} ~ M_{\odot} \, \rm{yr}^{-1}\,.
\end{equation}
An \emph{implicit} metallicity dependence exists as the evolution of the radius $R$ and $L$ depend on the mean molecular weight and hence $Z$, as mentioned earlier (e.g. Eq. \ref{MS_M_L}). 
Exceptions apply for very massive stars, e.g. luminous blue variables with luminosity $L>4000\,L_{\odot}$, where the mass-loss is modeled according to:
\begin{equation}
\dot{m} = 9.6 \times 10^{-15} \left( \frac{Z}{Z_{\odot}}\right) ^{0.5} R^{0.81}L^{1.24}m^{0.16} M_{\odot} \, {\rm yr}^{-1}\,.
\end{equation}
This is Eq. $2$ from \citet{Nieuw1990} but modified by the factor $Z^{0.5}$ \citep{Kudritzki1989}. 
Note that mass-loss can also occur as a result of mass transfer - having ultimately the same effect of moving a star along the MS towards lower effective temperature and hence lower luminosity.

\subsection{Stellar evolution of an entire population}\label{SSE_model}
\begin{figure*}
\centering
\begin{minipage}[t]{5.8cm}
\includegraphics[width=1\textwidth]{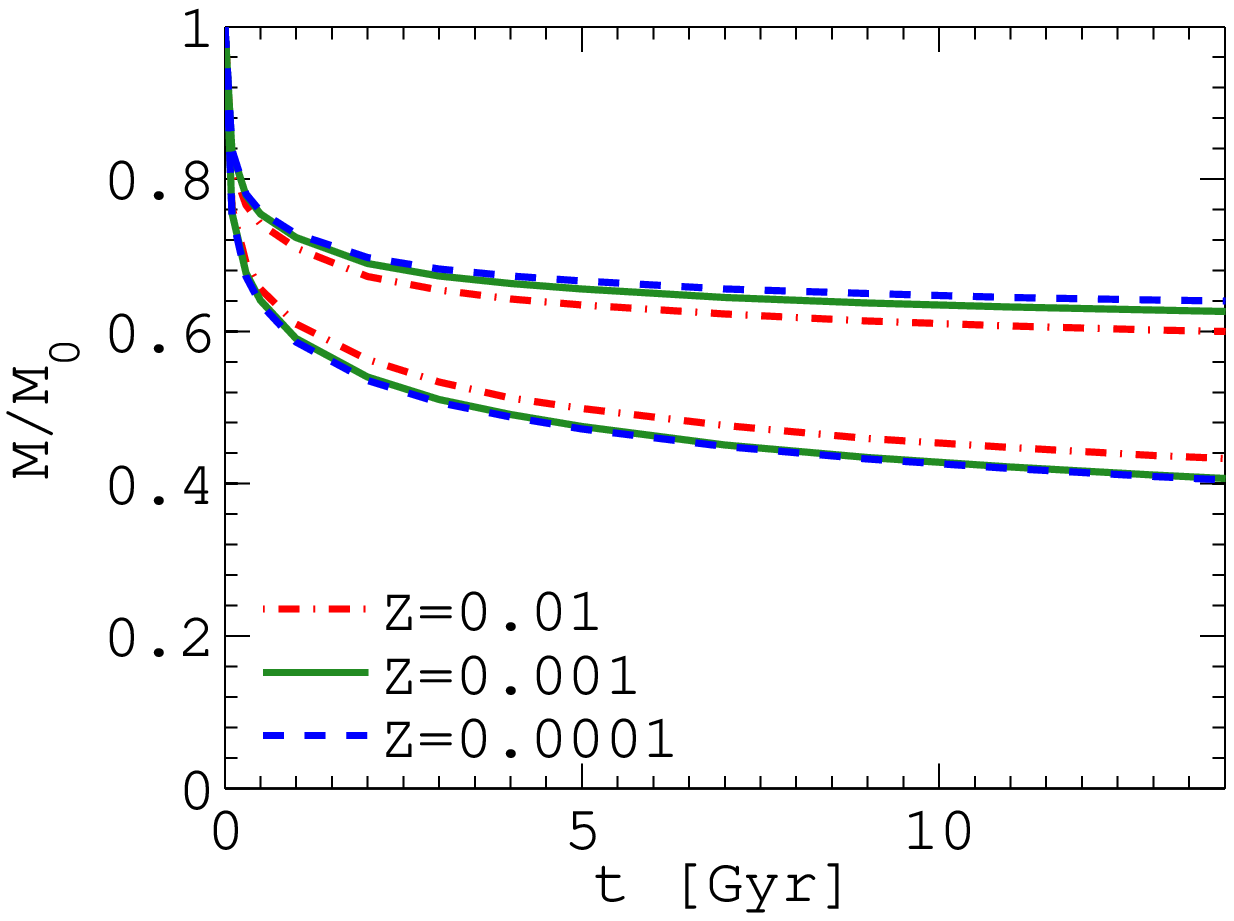}
\end{minipage}
\hspace{-0.1cm}
\begin{minipage}[t]{5.8cm}
\includegraphics[width=1\textwidth]{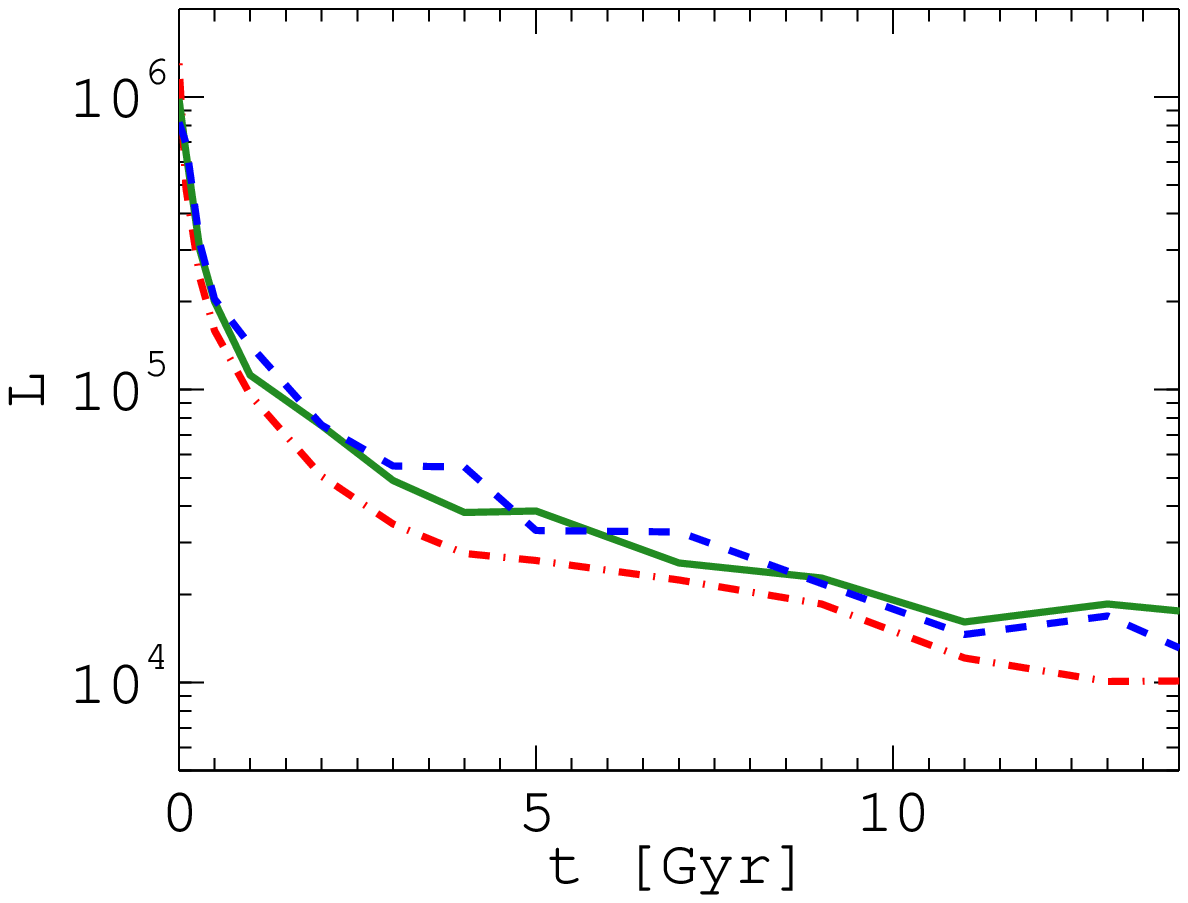}
\end{minipage}
\hspace{-0.1cm}
\begin{minipage}[t]{5.8cm}
\includegraphics[width=1\textwidth]{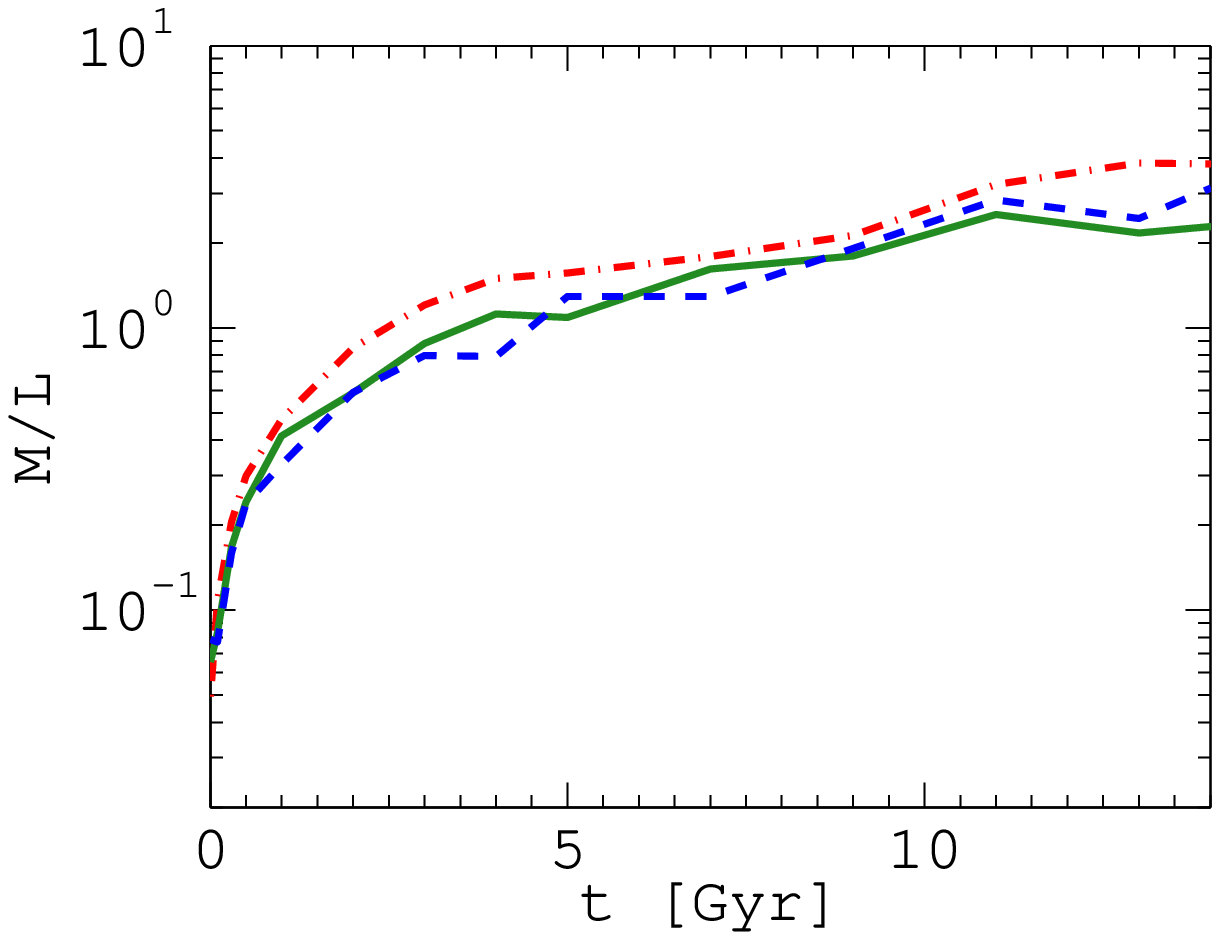}
\end{minipage}
\caption{Population of $105~000$ stars, corresponding to the set-up for our $N$-body models, evolved with only stellar evolution (i.e. no dynamical interactions). The metallicities are $Z=0.0001$ (dashed blue), $Z=0.001$ (solid green) and $Z=0.01$ (dashed-dotted red). \emph{Left panel:} The upper set of lines is the mass (scaled by the initial mass) of the \emph{entire} population of stars (including all remnants) while the lower set of lines only includes stars on the main sequence. This illustrates that the mass contained in MS stars is always higher for metal-rich clusters and that the overall mass evolution depends critically on the treatment of remnants. \emph{Middle panel:} Corresponding luminosity evolution (in units of $L_{\odot}$). In this case the treatment of remnants is not crucial. Note that even though the metal-rich model contains more stars on the MS, the metal-poor models have a higher luminosity. This is in agreement with Eq. \ref{MS_M_L} and implies that the higher luminosity of low-$Z$ stars is outweighing the fact that the metal-poor clusters have a lower MS turnoff mass at any given time than the metal-rich cluster (see Table \ref{tb:tb1}). Note that both metal-poor cases are expected to evolve in a similar fashion (also see Table \ref{tb:tb1}), however some variation is caused depending on the number of bright stars at any given time. \emph{Right panel:} Resulting mass to light ratio $M/L$ (in units of $M_{\odot}/L_{\odot}$). As expected from the luminosity evolution, the mass-to-light ratio is higher for the metal-rich model.}\label{fig:fig2}
\end{figure*}

To quantify the effects of stellar evolution on a non-dynamical population, we evolve $105\,000$ stars together through stellar evolution alone \citep{SSE}. This means that dynamical effects such as the influence of the galactic tidal field as well as the intrinsic $N$-body evolution within the cluster are ignored. The set-up of the initial masses of this population is identical to our $N$-body models introduced in Section \ref{method}, where the dynamical evolution is fully incorporated. In Fig. \ref{fig:fig2}, the mass, luminosity and mass-to-light ratio evolution of this model is illustrated for the three metallicities $Z=0.01$, $Z=0.001$ and $Z=0.0001$. At the Hubble time, $\approx 30\%$ of the initial stellar mass is lost purely due to stellar evolution and only $\approx 50\%$ of the initial mass in MS stars is still remaining (in agreement with \citealt{Baumgardt2003}). The overall mass of the low-$Z$ population stays higher throughout, while the mass contained in MS stars is always higher in the high-$Z$ population, as expected due to the higher MS turnoff mass. The luminosity (actually calculated as the V-band luminosity) drops by an order of magnitude within the first $\approx 2$ Gyr and roughly another magnitude over the next $10$ Gyr of evolution. We see that even though a high-$Z$ cluster will have a higher $m_{\rm{TO}}$, the luminosity of a metal-poor cluster remains $1.5-2$ times higher throughout the entire evolution - based on stellar evolution alone. This implies that the increased brightness of low-$Z$ stars is outweighing the higher number of MS stars in the high-$Z$ case.
As expected from the evolution of mass and luminosity in this non-dynamical model, the mass-to-light ratio $M/L$ is predicted to be higher by nearly a factor two for a metal-rich cluster. In a dynamically evolved model with a tidal field, the mass-to light ratios are likely to be modified as preferentially low-mass stars are lost from the outskirts of the cluster \citep{Baumgardt2003}. Low-mass MS stars are faint and have a high mass-to-light ratio. We will compare Fig. \ref{fig:fig2} to our dynamical models in Section \ref{ML_ratio}.

\subsection{Size: projection effects vs. internal dynamics}
For the MW, \citet{vdBergh1991} found that the GC half-light radius $r_{\rm{hl}}$ can be related to the galactocentric distance $d_{\rm{gc}}$ via $r_{\rm{hl}} \approx \sqrt{d_{\rm{gc}}}$ (see Fig. \ref{fig:fig1}). As the MW is the only galaxy where three-dimensional galactocentric distances are available, one has to rely on projected distances for extragalactic GC systems. Studies of extragalactic GC systems have shown that red and blue clusters are found to have different spatial distributions within the potential of their host galaxy: red clusters are distributed closer to the centre of the galaxy and subject to a stronger influence of the tidal field than the blue clusters \citep{BS2006}. A size difference ranging from $17\%$ \citep{Jordan2005} to $30\%$ \citep{Woodley2008} for the blue and red population has been found in numerous studies. Several scenarios have been proposed for the origin of this effect: projection effects and/or the effect of stellar evolution in combination with mass segregation, which we describe below. In addition, the possibility of different initial conditions during cluster formation have been proposed \citep{Harris2009} as well as different initial mass functions for metal-poor or metal-rich clusters \citep{Strader2009, Strader2011}.

\citet{Larsen2003} found that projection effects may account for the observed size difference of red and blue GCs, if the GC distribution flattens out near the centre of the galaxy (e.g. King profile) and there is a steep relation between cluster size $r$ and galactocentric distance $d_{\rm{gc}}$. However, this is not the case for either more centrally peaked distributions or shallower $r-d_{\rm{gc}}$ relations. \citet{Spitler2006} find in agreement with \citet{Larsen2003} that projection effects could explain the observed size difference in the Sombrero galaxy. A size gradient for GCs is found for small but not large (projected) galactocentric distances.

In contrast to this, \citet{Jordan2004} has found that the combined effects of mass segregation and MS lifetime lead to a size difference of low vs. high metallicity clusters. Under the assumption that the average half-mass radius does not depend on metallicity, the observed light-profiles were modeled with Michie-King multi-mass models and stellar isochrones leading to the result that a size difference of the observed magnitude arises naturally, with the metal-rich model having a half-light radius $\sim 14\%$ smaller than its metal-poor counterpart. The reasoning for this originates from the different speed in stellar evolution of stars with different $Z$ implying that the \emph{light} profile of a high $Z$ cluster can appear more concentrated. Unfortunately, in this approach the interplay between stellar dynamics and evolution was not considered. We note that \citet{Jordan2004} assumes the average half-mass radius to be independent of [Fe/H] - an assumption pointing to a universality in the formation and evolution of GCs. 

As part of the ACS Virgo Cluster Survey \citep{Cote2004} the sizes of thousands of globular clusters belonging to $100$ early type ellipticals in Virgo were measured \citep{Jordan2005}. They find in agreement with previous studies that the average half-light radius depends on the color of the GCs, with red GCs being $\approx 17 \%$ smaller than their blue counterparts. This size difference was proposed to originate from the effects of mass-segregation and metallicity, hence intrinsic cluster mechanisms as in \citet{Jordan2004}.

The arguments given above show that it is necessary to know the three dimensional galactocentric distance of GCs to their host galaxy to fully understand and disentangle the influence of the environment and metallicity on GC evolution. To be able to distinguish between those effects, we focus on metal-poor and metal-rich clusters at the same location, i.e. where both coexist. In the MW, $16$ GCs are located in the region between $7 \leq d_{\rm{gc}} \leq 10\,$kpc with a mean size of $r_{\rm{hl}}=3.95\,$pc. Four of those are metal-rich ([Fe/H]$>1.1$) and $12$ metal-poor. Thus we chose a galactocentric distance of $d_{\rm{gc}}=8.5\,$kpc for our models.


\section{SIMULATION METHOD \& CHOICE OF PARAMETERS}\label{method}
We use the direct $N$-body code \texttt{NBODY6} \citep{Aarseth1999, Aarseth2003} to construct and evolve our models. This state-of-the art $N$-body code takes advantage of the possibility to carry out such simulations on a graphics processing unit (GPU) coupled together with conventional central processing units \citep{Nitadori2012}. The simulations were carried out on Tesla S$1070$ graphics cards at Swinburne University. 

We use a Kroupa initial mass function (IMF: \citealt{Kroupa1993}) within the limits $0.1$ to $50\,M_{\odot}$ to populate our cluster model with stars. The beginning $t=0$ for the simulation corresponds to the zero-age MS and no gas is included in the models. The simulations start with $N_i=100\,000$ stellar systems, including a primordial binary frequency of $5\%$ (see Section \ref{bf}). These stars are initially distributed following a Plummer density profile
\begin{equation}
\rho(\mathbf{r})=\frac{3M}{4\pi R_{sc}^3}\left[1+\left(\frac{r}{R_{sc}}\right)^2\right]^{-5/2}
\end{equation}
\citep{Plummer1911, Aarseth1974} where $M$ is the cluster mass and $R_{sc}$ is a scale radius (see below). As the Plummer profile formally extends out to infinite radius, a cut-off at ten times the half-mass radius is applied  to avoid rare cases of stars at large distances \citep{Aarseth2003}. The individual initial positions and velocities are then assigned such that the cluster is in virial equilibrium. 

The cluster is subject to a constant, MW-like tidal field consisting of three components: a point-mass bulge, an extended smooth disc \citep{Miyamoto1975}, and a dark matter halo. We use $M_{\rm{b}}=1.5 \times 10^{10} \,M_{\odot}$ and $M_{\rm{d}}=5 \times 10^{10} \,M_{\odot}$ for bulge and disc mass, respectively \citep{Xue2008}. The scale parameters for the Miyamoto disc are $a=4\,$kpc (disc scale length) and $b=0.5\,$kpc (galactic thickness). Formally the disk extends to infinity but with this choice of parameters the strength has dropped to less than $0.1\%$ of the central value at a distance of $40\,$kpc. The dark matter halo follows a logarithmic profile $\Phi \propto v_0 ^2 \ln(d^2+b^2)^{0.5}$ \citep{Aarseth2003}. Here $d$ is the distance from the galactic centre at any given time, and $b$ is constrained such that the combined mass of the bulge, disk and halo give an orbital velocity of $v_0=220\,$km/s at a galactocentric distance of $d_{\rm{gc}}=8.5\,$kpc.

\begin{figure}
\centering
\includegraphics[width=0.47\textwidth]{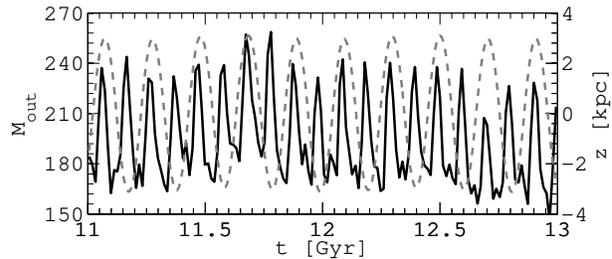}
\caption{Mass in the outskirts of the cluster (black solid line) and height $z$ above the galactic plane (dashed grey line). $M_{\rm{out}}$ is defined as the mass between one and two tidal radii. The galactic disc corresponds to $z=0$. Equivalent behaviour is observed for all models and metallicities. Approximately $30\,M_{\odot}$  are lost at every disc crossing.}\label{fig:fig3}
 \end{figure}

As mentioned earlier, we choose to place our clusters in an orbit at $d_{\rm{gc}} \approx 8.5\,$kpc to match an environment where red and blue clusters coexist within the MW (see Fig. \ref{fig:fig1}). The orbit is inclined $\approx 22 \deg$ to the galactic disc reaching a maximum height of $z \approx 3\,$kpc above the galactic plane. The apogalacticon is $8.8$ and perigalacticon $8.2\,$kpc with orbital period of $\approx 0.2\,$Gyr (see Fig. \ref{fig:fig3}). We chose a mid eccentricity to not start with extreme cases. The inclination results in a maximum $z=3\,$kpc, which is typical for many MW clusters \citep{Dauphole1996}.
During the lifetime of a cluster, stars are naturally lost due to dynamical relaxation, evolution and disc-shocking events. 
The tidal radius of a cluster in the Milky Way potential described above can be approximated as: 
\begin{equation}\label{rt_formula}
r_t \simeq \left( \frac{GM}{2 \Omega^2} \right) ^{1/3}
\end{equation}
\citep{Kuepper2010}, where $\Omega$ is the angular velocity of the cluster orbit and $G$ is the gravitational constant. Calculated at apogalacticon gives a $r_{\rm{t}}=52\,$pc, which we take as our initial value. This is adjusted as the cluster evolves according to the factor $M^{1/3}$. Stars are only removed from the cluster once their distance from the cluster centre exceeds twice the tidal radius. \citet{Gieles2011} have expressed the impact of the galactic tidal field on a cluster by quantifying a boundary $M_{\rm{lim}} < 10^5 M_{\odot} \times 4 \, \rm{kpc}/{R_{\rm{gc}}}$ below which clusters are tidally affected, whilst more massive clusters are tidally unaffected. The clusters in this study fall below this limit and hence are tidally limited.

Within the framework of \texttt{NBODY6}, the only remaining parameter is the scale radius $R_{\rm{sc}}$, which sets the initial cluster size or density and acts as a conversion factor between physical and $N$-body units. It is an ongoing debate as to how extended GCs are when they are born. Recently, it has been pointed out that GCs could be the remnants of much bigger stellar structures such as the nuclei of accreted dwarf galaxies \citep{Freeman1993, Boecker2008, Forbes2010}. In general, shortly after stellar nuclear fusion is ignited within a proto-cluster, the cluster it is expected to increase it's size as the remaining gas not incorporated into stars during star formation is ejected from the cluster. So far, globular cluster sizes at this early stage cannot be determined through observations. We choose $R_{\rm{sc}}=8$, corresponding to an initial three-dimensional half-mass radius of $r_{\rm{50\%}} \approx 6.2\,$pc. The half-mass radius evolves to $\approx 7\,$ pc at the Hubble time, but is a three dimensional quantity and hence a smaller half-mass radius by $25\%$ would be expected when measuring projected radii in two dimensions \citep{Fleck2006}. This places our models within the size range of observed clusters in the MW at $d_{\rm{gc}} \sim 8.5\,$kpc (see Fig. \ref{fig:fig1}) as well as in the large and small Magellanic Clouds \citep{Mackey2008}.

\begin{figure*}
\centering
\begin{minipage}[t]{5.8cm}
\includegraphics[width=1\textwidth]{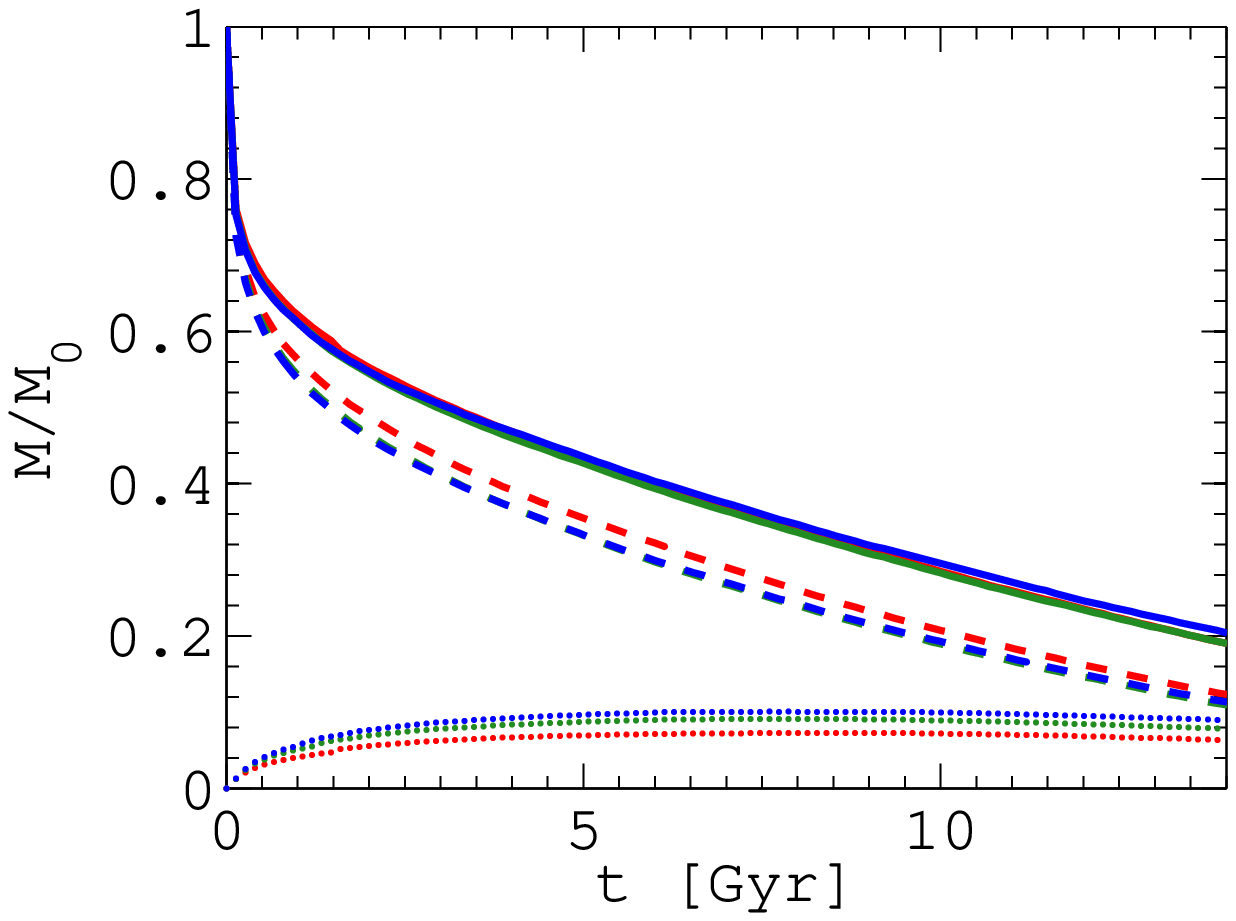}
\end{minipage}
\hspace{-0.1cm}
\begin{minipage}[t]{5.8cm}
\includegraphics[width=1\textwidth]{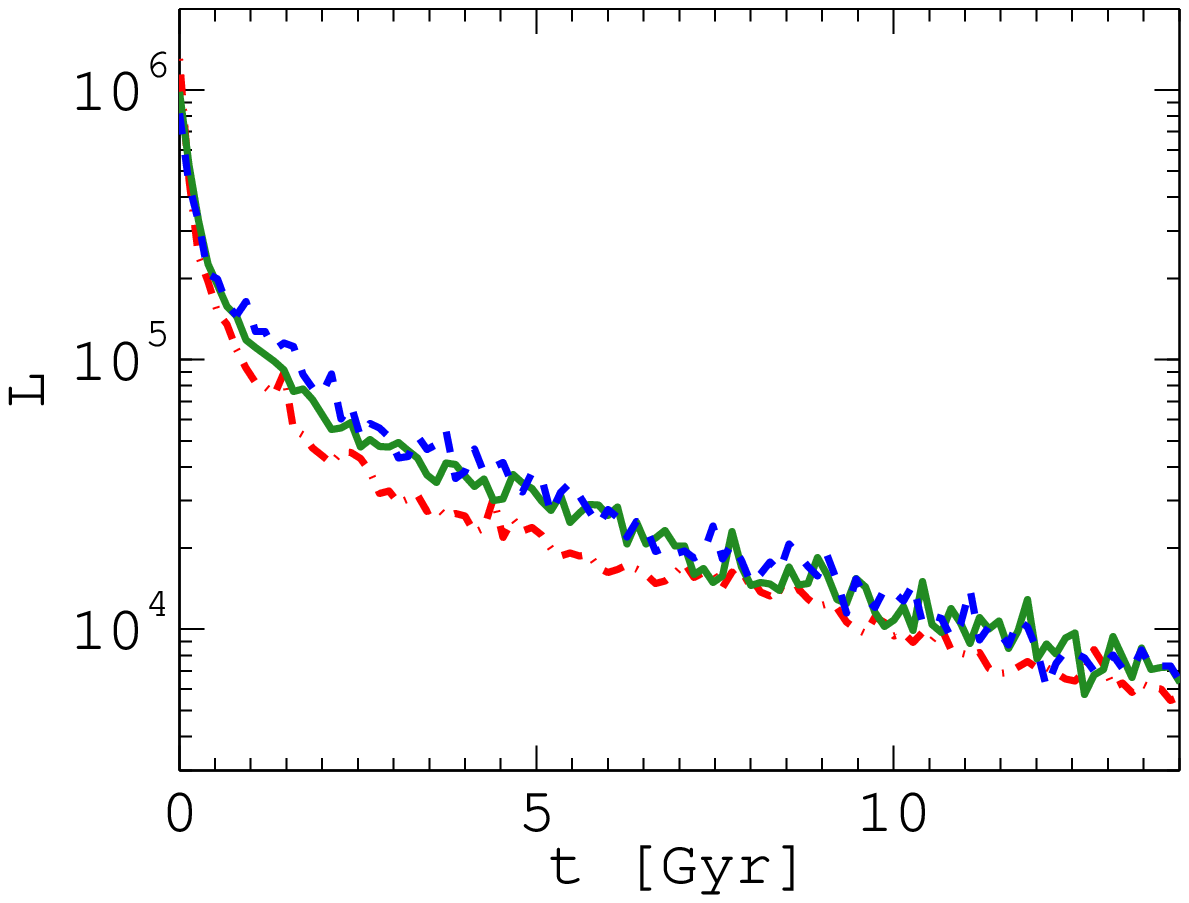}
\end{minipage}
\hspace{-0.1cm}
\begin{minipage}[t]{5.8cm}
\includegraphics[width=1\textwidth]{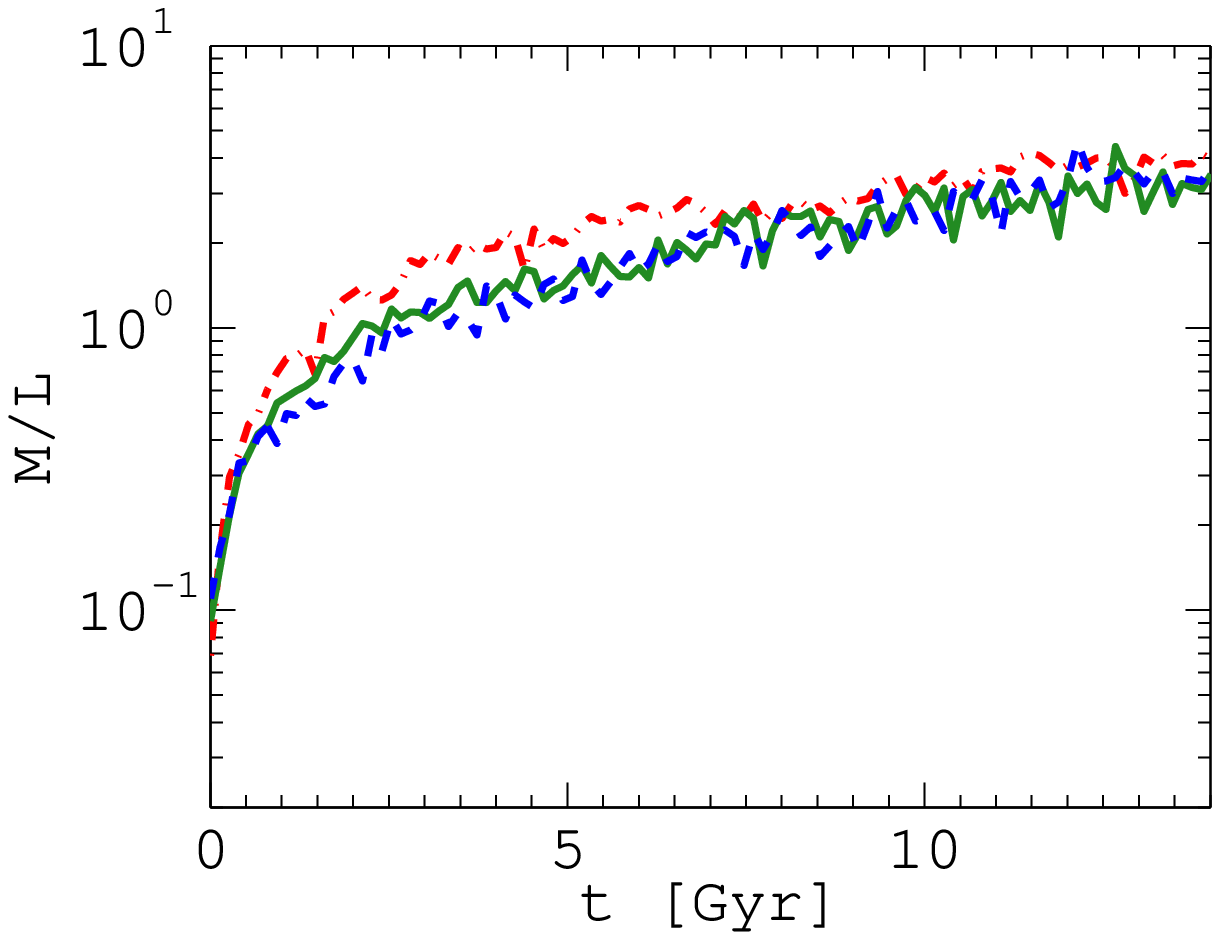}
\end{minipage}
\caption{Mass-loss rate for a cluster model dynamically evolved (including stellar evolution) - compared to Fig. \ref{fig:fig2} without dynamical interactions. \emph{Left Panel:} The solid line denotes the overall cluster mass (note that in contrast to Fig. \ref{fig:fig2} the high-Z population is no longer less massive than the low-Z populations). The dashed line is the mass contained in MS stars and the dotted line is the mass contribution from WDs. \emph{Middle panel:} Corresponding luminosity evolution. Although most apparent up to $\approx 7\,$Gyr, the metal-poor clusters stay more luminous throughout the entire evolution. As exected the overall luminosity is lower than in the non-dynamical case (Fig. \ref{fig:fig2}). \emph{Right panel:} Resulting mass-to-light ratio. Within the noise the values are equivalent to Fig. \ref{fig:fig2}. Hence the overall dynamical evolution has an impact on the mass of the cluster, but little effect on $M/L$. }\label{fig:fig4} 
\end{figure*}

\subsection{Binary fraction}\label{bf}
All models used in this study are evolved with the same number of initial stellar systems, $N_i=100\,000$, incorporating a binary fraction
\begin{equation}\label{binary_fraction}
b_f=\frac{N_b}{N_s+N_b} = 0.05\,.
\end{equation}
This translates into $N_s=95\,000$ single stars and $N_b=5000$ binary systems, therefore $105\,000$ stars in total. Some of these primordial binary systems may be disrupted early on, while new binaries form during the cluster evolution due to two- or three-body interactions. Open cluster studies found in the literature are usually evolved with binary fractions of $0.2-0.5$ \citep{Hurley2004, Hurley2005, Trenti2007}, as observations find higher binary fractions in these clusters (e.g. \citealt{Montgomery1993, Richer1998} for M$67$). Much lower binary fractions are observed in GCs: \citet{Milone2012} have measured the binary fractions of $59$ GCs in the MW and commonly find values around $b_f \approx 0.05$. Binary systems in GC models have proven to be important from a dynamical point of view: even a small binary fraction in the core can be sufficient to heat the cluster core enough to postpone core-collapse significantly \citep{Hut1992, Heggie2006}. 

Within \texttt{NBODY6}, standard binary evolution is treated according to the binary algorithm of \citet{BSE} where circularization of eccentric orbits as well as angular momentum loss mechanisms are modeled. Wind accretion from one binary component to the other is possible as well as mass transfer when either star fills its Roche lobe. Stable hierarchical three- and four-body systems are detected and evolved \citep{Mardling2001}, with single-binary and binary-binary encounters followed directly. This allows for the replacement of one member of a binary by an incoming star, formation of binaries in few-body encounters and direct collisions \citep{Kochanek1992}, often leading to the formation of exotic stars such as blue stragglers. Nearby stars can perturb binary systems and cause chaotic orbits \citep{Mardling2001}.

\subsection{Treatment of remnants}\label{tr_remn}
Neutron stars are assumed to be subject to a velocity kick arising from asymmetries during their formation through core-collapse supernovae, with observations of NSs indicating a vast range of velocities from several up to hundreds of km/s. Such velocities are easily in excess of a typical GC escape velocity and, in combination with observations of substantial NS populations in GCs, is known as the neutron-star retention problem \citep{Pfahl2002}. Indeed, X-ray sources (e.g. \citealt{Woodley2008} in the case of NGC 5128) and milli-second pulsars (\citealt{Bogdanov2011} in the core of NGC 6626) indicate that NSs and BHs are common and even BH-BH binaries may exist. 
In $N$-body simulations, several different methods to assign velocity kicks to NS or BH remnants have been used in the past. \citet{Baumgardt2003} simply retain all NSs. With their IMF not reaching masses higher than $15\,M_{\odot}$, the number of NSs is not excessive and no BHs form. \citet{Mackey2007} retain all stellar-mass remnant BHs whilst using an IMF up to $100\,M_{\odot}$. In contrast to this, \citet{Hasani2010} retain no NSs or BHs. \cite{Hurley2010} use a Gaussian velocity kick distribution peaked at $\approx 190\,$km/s for both NSs and BHs, where the formation of a BH-BH binary is later observed to postpone core-collapse. 

In this study, we adopt an intermediate approach by choosing $v_k$ at random from a flat kick distribution in the range $0-100\,$km/s and assigning this to NSs and BHs at their birth. Because of the low escape velocity $v_e=\sqrt{2GM/r} \approx 4.7\,$km/s at the half-mass radius, or $v_e \approx 2.8\,$km/s at the tidal radius (both at a cluster age of $500\,$Myr), this reproduces a retention fraction of $\approx 5\%$ \citep{Pfahl2002}. We use the same algorithm to assign kick velocities to BHs at their formation \citep{Repetto2012}. We note that the metallicity influences the mass of the remnants. In our model, the maximum BH mass is $\approx 30~M_{\odot}$ for metal-poor progenitors and $\approx 10\,M_{\odot}$ for their metal-rich counterparts \citep{SSE, Belczynski2010}.

\subsection{MODELS}\label{models}
We evolve three sets of models a), b) and c) with identical set-up apart from the random number seed for the initial particle distributions. Each set consists of three models with metallicities $Z=0.0001$, $Z=0.001$ and $Z=0.01$ (see Table \ref{tb:tb2}), i.e. low, intermediate and high metallicity. GCs in the MW are found within the metallicity range $-2.37 \leq$ [Fe/H] $\leq 0$ (\citealt{Harris1996}). The intermediate metallicity case $Z=0.001$ of this study already corresponds to a metal-poor cluster in the MW (and also other galaxies). The low-$Z$ case $Z=0.0001$ is an example from the metal-poor end of the metallicity distribution. We expect these two low-metallicity clusters to exhibit similar evolution to each other (e.g. the MS turnoff masses agree fairly well: see Table \ref{tb:tb1}) but distinct from the high-metallicity case. This has previously been noted by \citet{Hurley2004}. All models are evolved up to $14\,$Gyr, while we concentrate our analysis at typical GC age of $12$ Gyr \citep{Hansen2007}. 
\begin{table*}
 \centering
  \caption{Metallicities and initial masses for all models at $t=0$ and various parameters at $12$ Gyr: mass $M$, number of stars $N$, binary fraction $b_f$ (Eq. \ref{binary_fraction}), number of MS stars $N_{\rm{MS}}$ as well as the total mass contained in MS stars $M_{\rm{MS}}$, mass locked in WDs $M_{\rm{WD}}$ and the number of NSs and BHs, all at $12\,$Gyr. The variation in the initial cluster mass arises from the difference in random seed when drawing stars from the IMF. Note the consistently higher WD mass for metal-poor clusters: WD masses are higher for same-mass progenitors and in addition more stars have already turned off the MS into WDs.}\label{tb:tb2}
  \begin{tabular}{@{}rrcccccccccc@{}}
    & Z & [Fe/H] & $M_{0}$ & $M$ ($12$ Gyr) & $N$ & $b_f$ & $N_{\rm{MS}}$ & $M_{\rm{MS}}$ & $M_{\rm{WD}}$ & $N_{\rm{NS}}$ & $N_{\rm{BH}}$\\ \hline \hline
   a) &  $0.0001$ &  $-2.3$ &  $6.43 \times 10^4$ & $1.57 \times 10^4$ & $35699$ & $0.0597$ & $27626$ & $9.48 \times 10^3$ & $6.14 \times 10^3$ & $18$ & $2$ \\
       &  $0.001$   &  $-1.3$ &  $6.43 \times 10^4$ & $1.51 \times 10^4$ & $34787$ & $0.0595$ & $27025$ & $9.40 \times 10^3$ & $5.51 \times 10^3$ & $31$ & $2$ \\
       &  $0.01$     &  $-0.3$ &  $6.43 \times 10^4$ & $1.45 \times 10^4$ & $35680$ & $0.0604$ & $27975$ & $1.03 \times 10^4$ & $4.44 \times 10^3$ & $24$ & $2$ \\ \hline
       
   b) &  $0.0001$ &  $-2.3$ &  $6.42 \times 10^4$ & $1.56 \times 10^4$ & $35958$ & $0.0606$ & $27948$ & $9.57 \times 10^3$ & $6.10 \times 10^3$ & $25$ & $2$ \\
       &  $0.001$   &  $-1.3$ &  $6.42 \times 10^4$ & $1.50 \times 10^4$ & $34654$ & $0.0595$ & $27036$ & $9.39 \times 10^3$ & $5.42 \times 10^3$ & $23$ & $3$ \\
       &  $0.01$     &  $-0.3$ &  $6.42 \times 10^4$ & $1.51 \times 10^4$ & $35017$ & $0.0615$ & $28229$ & $1.04 \times 10^4$ & $4.38 \times 10^3$ & $16$ & $3$ \\ \hline
       
   c) &  $0.0001$ &  $-2.3$ &  $6.36 \times 10^4$ & $1.59 \times 10^4$ & $36149$ & $0.0585$ & $28065$ & $9.66 \times 10^3$ & $6.13 \times 10^3$ & $26$ & $0$ \\
       &  $0.001$   &  $-1.3$ &  $6.36 \times 10^4$ & $1.52 \times 10^4$ & $34804$ & $0.0606$ & $27015$ & $9.43 \times 10^3$ & $5.50 \times 10^3$ & $24$ & $2$ \\
       &  $0.01$     &  $-0.3$ &  $6.36 \times 10^4$ & $1.49 \times 10^4$ & $34596$ & $0.0628$ & $27896$ & $1.03 \times 10^4$ & $4.32 \times 10^3$ & $27$ & $1$ \\
 \end{tabular}
 \end{table*}
 %
 \section{EVOLUTION}\label{evolution}
During cluster evolution, stars are lost in three ways: i) an increase of velocity during two-body encounters (evaporation) or ejection following three or four-body encounters, ii) velocity kicks owing to SN explosions and iii) tidal stripping and disc shocking (i.e. the influence of the tidal field of the host galaxy). These three effects cannot be completely disentangled: the former two might bring a star close to or even beyond the tidal boundary $r_t$ (Eq. \ref{rt_formula}) such that when crossing the galactic disc, stars are easily removed from the system. The periodicity of this event is $\approx 100\,$Myr and causes the number of stars within the cluster envelope between one and two tidal radii to continually fluctuate between  $180-240\,M_{\odot}$, with $\approx 30\,M_{\odot}$ lost each time the disc is crossed (see Fig. \ref{fig:fig3}). The evolution of a star cluster is linked to the two-body relaxation time:
\begin{equation}
t_{\rm{rh}}=\frac{0.14 N}{\ln(0.4N)} \sqrt{\frac{r_{\rm{50\%}}^3}{G\,M}}
\end{equation}
(\citealt{Spitzer1971, BinneyTremaine2008}, see also \citealt{Hurley2001}). For our models, the relaxation time is highest at $\approx 2.5$ Gyr when $t_{\rm{rh}}$ corresponds approximately to the cluster lifetime at that point. The relaxation time then decreases to roughly one Gyr at $12$ Gyr of cluster age. The half-mass relaxation timescale is not significantly affected by the metallicity: variations up to $10\%$ occur. This means that the clusters of different metallicity are dynamically of similar age, which is not the case from a stellar evolution point of view. As can be seen in Table \ref{tb:tb2}, all three metallicity models have similar mass and number of stars at $12$ Gyr, whereas the distribution of mass among MS stars and remnants differs (the metal-poor cluster containing more mass in remnants).

\begin{figure}
\centering
\includegraphics[width=0.47\textwidth]{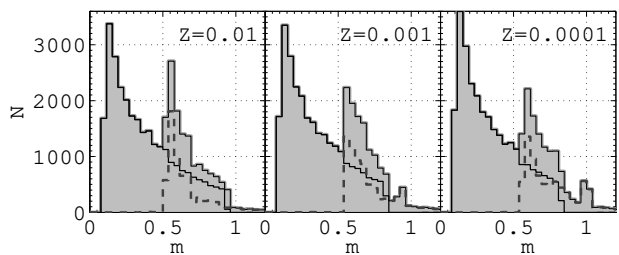}
\caption{Mass function of the dynamically evolved stellar population at $12$ Gyr for $Z=0.01$ (left panel), $Z=0.001$ (middle panel) and $Z=0.0001$ (right panel). Here we focus on model set b) but the behaviour is similar for all sets. The grey area is the entire population of stars, the thin black line the remaining stars on the main sequence and the dashed line the contribution of white dwarfs (peaked at $\approx 0.6~M_{\odot}$). For stars with $M \le 0.5M_{\odot}$ the population is made entirely out of MS stars. For metal-poor clusters, the MS turnoff is noticeably smaller (see also Table \ref{tb:tb1}). The number of NSs and BHs is insignificant compared to MS stars and WDs.}\label{fig:fig5}
\end{figure}

Fig. \ref{fig:fig4} is a reproduction of Fig. \ref{fig:fig2}, but now for our full $N$-body models. The three panels are again mass, luminosity and $M/L$. In Fig. \ref{fig:fig2} we only considered the mass-loss owing to stellar evolution, causing $\approx 40\%$ mass-loss up to $12\,$Gyr. For Fig. \ref{fig:fig4} the dynamical interactions are taken into account, resulting in an additional mass-loss of the same order, leaving a cluster mass of $\approx 25\%$ after $12\,$Gyr. Hence the three effects i)-iii) mentioned above are together responsible for approximately half the mass-loss of the cluster, while stellar evolution alone is responsible for the rest of the mass-loss. It can also be seen in Fig. \ref{fig:fig4} that nearly $40\%$ of the mass at $12\,$Gyr is contained in WDs. All stars are split into their relevant stellar populations in Fig. \ref{fig:fig5} for further illustration, also at a cluster age of $12\,$Gyr. While stars below $0.5\,M_{\odot}$ exclusively are on the main sequence, the contribution of WDs is significant for higher masses, causing a second peak in the mass function at $\approx 0.6\,M_{\odot}$.

There is always more (luminous) mass contained in MS stars in the metal-rich cluster, which is expected from the higher MS turnoff mass (see Table \ref{tb:tb1} and Fig. \ref{fig:fig5}), while the low-$Z$ clusters are more luminous in spite of this. See Section \ref{ML_ratio} for further details of the evolution of luminosity and mass-to-light ratio. Even though the MS turnoff is higher for metal-rich clusters, the number of MS stars is not always the highest (see Table \ref{tb:tb2}, column $7$). The fluctuation of $N_{\rm{MS}}$ is mainly due to the fact that the number of low-mass stars with $m\leq 0.2\,M_{\odot}$ varies depending on metallicity: the mass in those stars is typically $5-10\%$ higher for $Z=0.0001$ than for $Z=0.001$ or $Z=0.01$. While statistical noise may be responsible for some of the fluctuations, the lowest-$Z$ cluster also has the highest mass and hence a slightly higher escape velocity.

\subsection{Binary systems}
The binary fraction of initially $0.05$ slightly increases to $\approx 0.06$ at $12\,$Gyr for any metallicity or model (see Table \ref{tb:tb2}), where the binary fraction for the high-Z model is always slightly higher than for low metallicities. While some of the initial systems may easily disrupt, others form during few-body encounters. Hard binaries \citep{Heggie1975, Hut1992} have been shown to successfully halt core-collapse over large periods of time and BH binary systems in particular can heat the core substantially. With an initial mass function up to $50\,M_{\odot}$ and the inclusion of stellar evolution, black hole remnants will form early on in the cluster evolution. While most BHs get ejected almost immediately (e.g. Section \ref{tr_remn}), remaining BHs will sink towards the centre of the cluster owing to mass segregation. While doing so, they may become part of a binary or triple system, breaking up a previously existing binary system. Once BHs are part of binary systems, BH-BH binaries can easily form in a further encounter through exchange interactions. All BH-BH binaries in this stydy are dynamical binaries, having formed through such few-body interactions. This also means each component in a BH binary is the remnant of a high-mass MS progenitor that was either a single star or born in a binary system that later disrupted. 

\begin{figure*}
\centering
\includegraphics[width=0.9\textwidth]{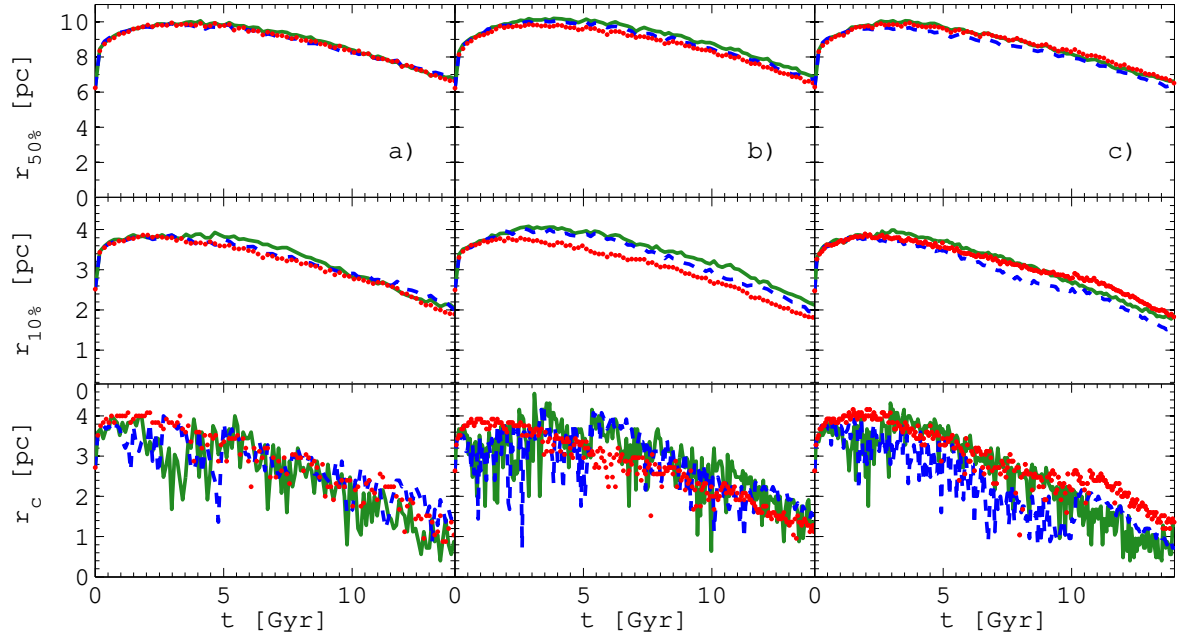}
\caption{Evolution of lagrangian radii and core radius with models a) on the left, b) in the middle panel and c) on the right. \emph{Top:} half-mass radius. Slight size differences between the models occur, however this is not primarely related to metallicity. Size differences are originating from few-body encounters and high-energy binary systems in the centre of the cluster and are enhanced in the $10\%$ lagragian radius (\emph{middle}). \emph{Bottom:} The $N$-body core radius $r_{\rm{c}}$. Short-term effects on the core radius are often linked to high-energetic binaries or the presence of BHs in the core, which can severely impact the evolution of $r_{\rm{c}}$. }\label{fig:fig6}
\end{figure*}

\subsection{Cluster size}
Owing to the cumulative effects of mass-loss, two-body relaxation and the influence of the tidal field, the models are expected to go through an initial expansion, followed by contraction. In Fig. \ref{fig:fig6} this is shown by means of the three-dimensional half-mass radius $r_{\rm{50\%}}$. We find no metallicity dependence on the half-mass radius. Moving further inwards, we look at the $10\%$ Lagrangian radius $r_{10\%}$ (middle panels of Fig. \ref{fig:fig6}) and the $N$-body core radius $r_{\rm{c}}$ (bottom panel). Small differences in $r_{10\%}$ are evident for the different models, noting that this inner radius is susceptible to the actions of highly  energetic binaries in the core, even so, the evolution of $r_{\rm{10\%}}$ remains fairly steady. The $N$-body core radius $r_{\rm{c}}$ is similar in size to the $10\%$ Lagrangian radius, however we see that $r_{\rm{c}}$ is heavily fluctuating when BHs, BH-BH binaries or otherwise energetic binary systems are present (all of which are more likely to reside in the central regions owing to mass segregation). 

The $N$-body core radius is not to be confused with an observational King-core radius, as the $N$-body core radius is a density weighted mean distance to the cluster centre (not taking luminosity into account). In the procedure of calculating $r_{\rm{c}}$, the mean density (in terms of mass) of the six neighboring stars is calculated for each star \citep{Casertano1985}, introducing a large bias towards stars in the neighborhood of BHs: a BH might be up to $28\,M_{\odot}$, a binary BH up to twice as much, while a MS is less massive than two solar masses after one Gyr of cluster evolution. We note that the $N$-body core radius is consistently less fluctuating at high metallicity than in the lower metallicity cases. This is not a sampling effect.  Instead, it results from remnant masses being lower for the high-$Z$ population. With BH masses only up to $10\,M_{\odot}$, the density contrast around stars will be less steep. BH-BH binaries can mimic core-collapse (Fig. \ref{fig:fig6}) when indeed just a subsystem of stars is responsible for this effect.

In addition, peaks in the core radius can (but don't have to be) closely correlated with highly-energetic binary systems. As an example, the drop in $r_{\rm{c}}$ for the low-$Z$ model b) in Figure \ref{fig:fig6} (middle panel) at $2\,$Gyr is caused by a short-period binary composed of two carbon-oxygen white dwarfs of masses $0.7$ and $0.8~M_{\odot}$.  At $t=1.85$ Gyr, the two WDs merged and the product was subsequently ejected from the cluster. The maximum binding energy before coalescence is $141\,M^2_{\odot}/AU$. This is followed by another dip in $r_{\rm{c}}$ at $2.6\,$Gyr, when the core radius shrinks to $0.72$ pc. At this point, more than  $50\%$ of the core-mass is contained in BHs and a BH-BH binary forms.

In the low-$Z$ model of set a), the $N$-body core radius drops by more than factor of two to $1.4$ pc at $5\,$Gyr. This is caused by a chain of reactions involving four remnant BHs (out of ten present at that time). The masses of the four BHs are $27$, $26$, $14$ and $11\,M_{\odot}$, respectively. Initially, the least massive BH is ejected from this four-body subsystem, and leaves the cluster. The remaining three form a short-lived triple-system which ends with a BH-BH binary and a single BH being ejected from the core as a result from enhanced velocities obtained in the interaction. This implies that four of the most massive components are lost from the core within a time frame of only $40\,$Myr.

\begin{figure*}
\centering
\includegraphics[width=0.8\textwidth]{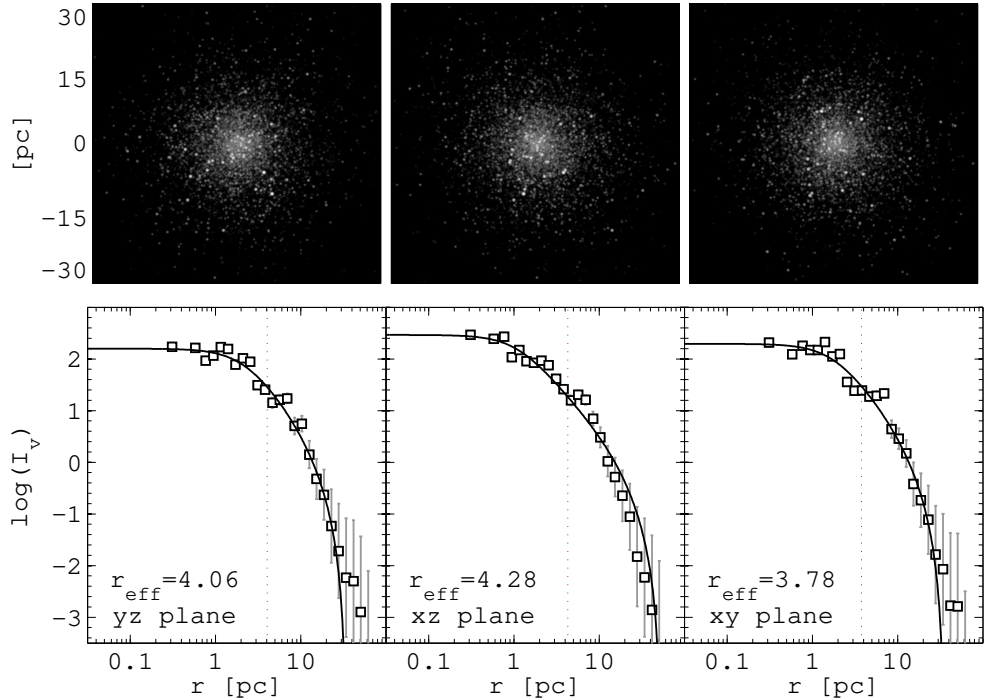}
\vspace{-1cm}
\caption{Example fits for a cluster at the age of $13$ Gyr. The three panels denote the same cluster at the same time, projected along the $x-$, $y-$ and $z-$ axis. The corresponding snapshot is printed above. Each snapshot is fitted individually. The measured data points for the surface brightness profile are denoted by black sqares with poisson error, the black line is the \texttt{gridfit} King$66$ fit. The resulting effective radius $r_{\rm{eff}}$ is indicated by the red dotted line.}\label{fig:fig7}
 \end{figure*}

We conclude that the metallicity has \emph{no} effect on the half-mass radius or other scaling parameters based on cluster mass. However as Figs. \ref{fig:fig2} and \ref{fig:fig4} already indicate - the metallicity influences the overall luminosity of GCs with high-$Z$ clusters being fainter than metal-poor clusters. To explore this possibility in more detail, we measure the half-light of effective radius $r_{\rm{eff}}$ by fitting \citet{King1966} models to our clusters - analogous to sizes are measured from observations. We illustrate this method in Section \ref{surf_br}. 

\subsection{Surface brightness and half light radii}\label{surf_br}
Among other properties, the output of \texttt{NBODY6} incorporates the mass, luminosity and radius for each star. This means effective temperatures can easily be calculated and this data can be cross convolved with stellar atmosphere model calculations \citep{Kurucz1979} to obtain Johnson V-band magnitudes. 

\begin{figure*}
\centering
\includegraphics[width=0.8\textwidth]{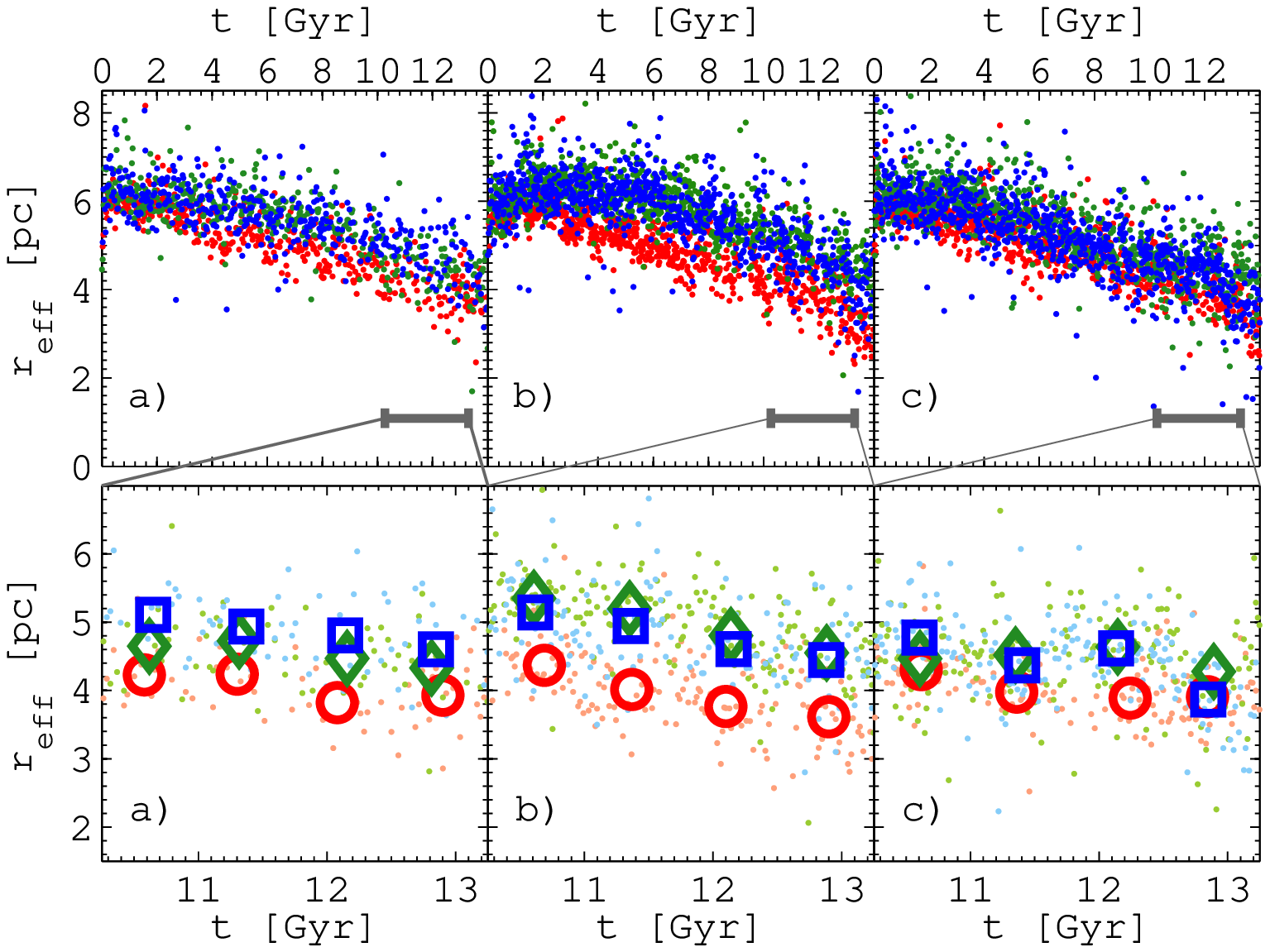}
\caption{Half-light or effective radius $r_{\rm{eff}}$ from \citet{King1966} model fits using \texttt{gridfit} \citep{McLaughlin2008}. In the top panels, the overall evolution of the half-light radius is indicated for all sets of models: a) on the left, b) in the middle and c) on the right. Of greatest interest is the data at late times, which are highlighted below. Average cluster sizes for each metallicity are calculated for the intervalls $10.25-11$ Gyr, $11-11.75$ Gyr, $11.75-12.5$ Gyr and $12.5-13.25$ Gyr, using blue squares for the low-$Z$, green diamonds for intermediate and red circles for the high-$Z$ case. It is clearly seen, that the metal-poor cluster snapshots (blue) have a larger observed half-light radius than the metal-rich (red) snapshots. The average sizes are summarized in Table \ref{tb:tb3}.}\label{fig:fig8}
 \end{figure*}

We project this data in a two dimensional image and slightly smooth it with a Gaussian filter (see Fig. \ref{fig:fig7} for an example of a cluster at the age of $13\,$Gyr). This means the light of each star is  conserved, but not contained within one single pixel, which implies that the starlight can be divided between consecutive bins when creating a surface brightness profile, which is crucial in cases of very bright stars. For each model, at each snapshot three such images are obtained by using the degree of freedom to project in either the $x$, $y$ or $z$ direction (in theory multiple projections are possible, see \citealt{Noyola2011}) and a surface brightness profile is obtained separately for each projected snapshot (Fig. \ref{fig:fig7}). For simplicity, we assume a background of zero.
We chose to fit \citet{King1966} models as they have shown to be a robust solution to fit GCs. Another option would be \citet{Wilson1975} models, having a greater sensitivity in the outer regions of the cluster \citep{McLaughlin2008}. However, in this work we are not investigating tidal fluctuations but the overall cluster evolution, which the King models are well suited for. Since there is no analytical solution for the surface density of this model, a grid of model fits has to be pre-calculated. We utilize the \texttt{gridfit} code \citep{McLaughlin2008} where this has been done. Each snapshot is fitted three times according to the three different projections along the $x$-, $y$- and $z$-axes, as illustrated in Fig. \ref{fig:fig7}. Obvious bad fits are rejected from further analysis (note that no bright stars have been masked for fitting). For each given time, the final effective radius is the mean along all three projections.

The result is plotted in Fig. \ref{fig:fig8} over the entire evolution of the cluster. Similar to the half-mass radius, an initial expansion when mass-loss is dominated by stellar evolution winds from massive stars in the core is followed by a contraction when the mass-loss is dominated from the cluster boundary. Yet there are differences in comparison to the half-mass radius: Firstly, $r_{\rm{eff}}$ is approximately half as large as the half-mass radius. As $r_{\rm{50\%}}$ is a three dimensional quantity, $r_{\rm{eff}}$ is expected to be only $3/4$ as large simply due to projection effects. A size difference further to this implies that the luminosity alters the measured cluster size. Secondly, there is a clear effect of the metallicity on the $r_{\rm{eff}}$ evolution of the clusters: the metal-poor clusters are consistently observed to be larger than their metal-rich counterparts. 

Also in Fig. \ref{fig:fig8} we highlight the time window of $10-13\,$Gyr which is of most significance for old GCs. The data is averaged over $\delta t = 750\,$Myr windows: $t_{10}=10.25-11$ Gyr, $t_{11}=11-11.75\,$Gyr, $t_{12}=11.75-12.5\,$Gyr and $t_{13}=12.5-13.25\,$Gyr. The results are summarized in Table \ref{tb:tb3} and combined give an overall size difference of $\approx 17\%$ between red and blue GCs. If split into sets, the difference is $19$, $21$ and $10\%$ for sets a), b) and c), respectively. This result implies that the observed size difference between the metal-poor and metal-rich GC sub-populations can (at least partly) be explained by the effects of metallicity. 

\begin{table}
\centering
\caption{Average cluster sizes measured for \emph{all} sets for the intervals $t_{10}=10.25-11\,$Gyr, $t_{11}=11-11.75\,$Gyr, $t_{12}=11.75-12.5\,$Gyr and $t_{13}=12.5-13.25\,$Gyr. In the bottom line the size difference $\Delta r = r_{\rm{b}}-r_{\rm{r}}$ is given for the corresponding time interval, where $r_{\rm{b}}$ is the average cluster size observed for blue, metal-poor and $r_r$ for red, metal-rich clusters. The overall size difference for all ages is $17\%$.}\label{tb:tb3}
\begin{tabular}{ l  c c c c}
                    & $t_{10}$       & $t_{11}$        & $t_{12}$          &  $t_{13}$ \\ \hline
$Z=0.01$     & $4.30$ pc    & $4.08$ pc     &  $3.85$  pc     &  $3.82$ pc \\
$Z=0.001$   & $4.82$ pc    & $4.81$ pc     &  $4.61$  pc     &  $4.39$ pc \\
$Z=0.0001$ & $5.01$ pc    & $4.75$ pc     &  $4.64$  pc     &  $4.31$ pc \\ \hline
$\Delta ~r$    &  $16.5\%$     &  $16.4\%$     & $20.5\%$        & $12.6\%$  \\
\end{tabular}
\end{table}

\subsection{Origin of the size difference and influence of remnants}
\begin{figure*}
\centering
\begin{minipage}[t]{8.5cm}
\includegraphics[width=1\textwidth]{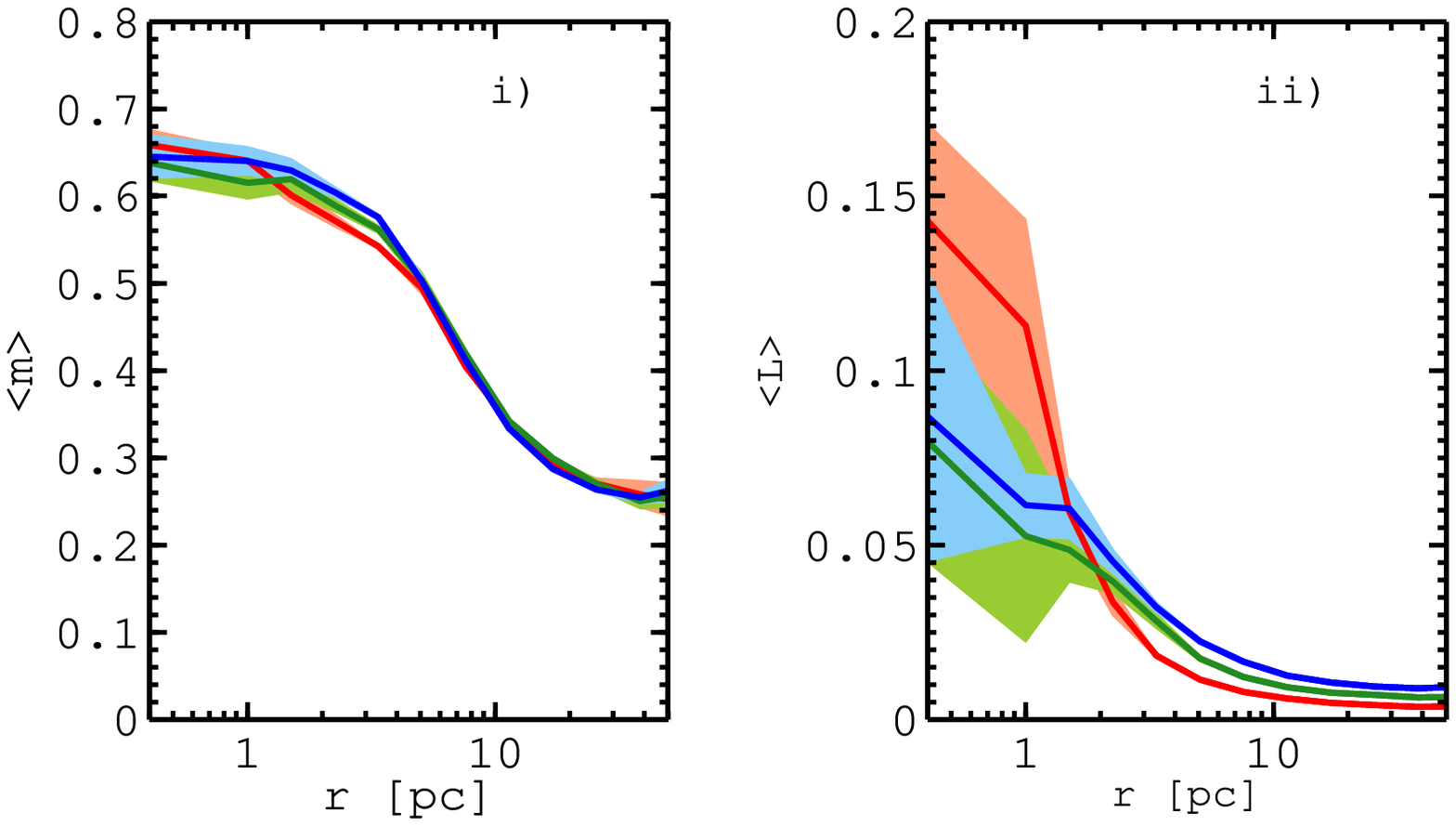}
\end{minipage}
\hspace{-0.1cm}
\begin{minipage}[t]{8.5cm}
\includegraphics[width=1\textwidth]{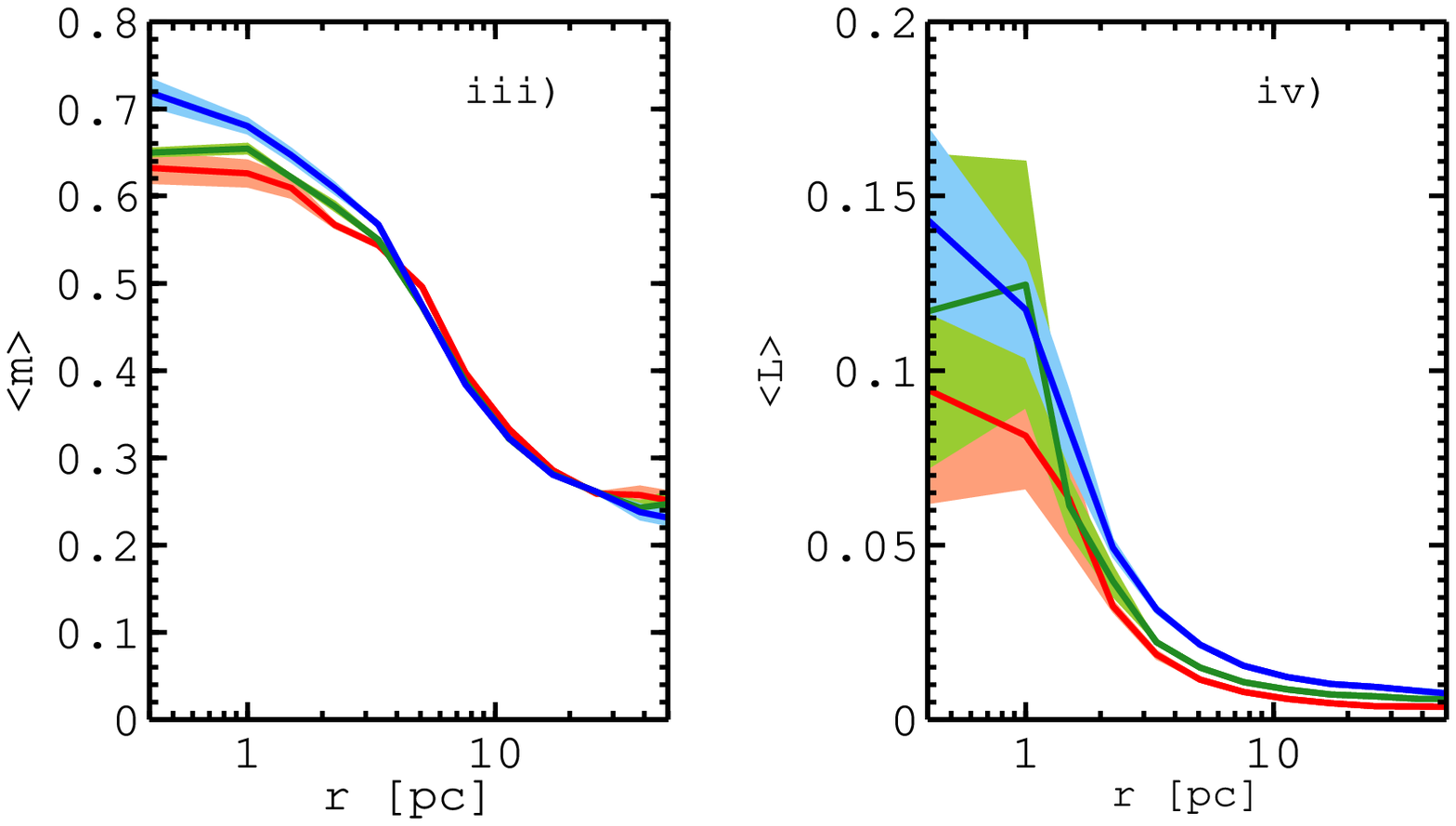}
\end{minipage}
\caption{Radial profiles of the average stellar mass (panel i) and average luminosity (panel ii)) for different metallicities. This is for model b) at an age of $\approx11.5\,$Gyr, averaged over ten consecutive snapshots (covering about $130\,$Myr). The shaded regions indicate the errors involved, calculated as the standard deviation from the mean within those ten snapshots. In calculating the average mass all stars and remnants are taken into account, while for the average luminosity only stars not yet in the remnant phase are taken into account (e.g. only luminous stars). There is a general trend for the luminosity distribution to be steeper in the high metallicity case. However beyond the core of the cluster, the metal-poor cluster has a higher average luminosity. 
The panels iii) and iv) are a repeat of the panels on the left, but for a set of models without NS or BH remnants. While the overall evolution for these models is similar to the other models in this study, we do not observe a significant size difference.}\label{fig:fig9}
 \end{figure*}

\begin{figure}
\centering
\includegraphics[width=0.45\textwidth]{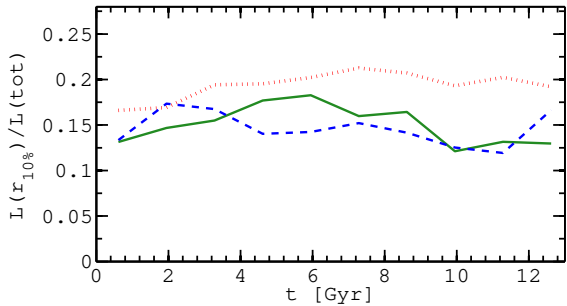}
\caption{Evolution of the luminosity contained withing the $10\%$ Lagrangian radius normalized by the total luminosity at that time, for model set b). 
The high-$Z$ cluster (dotted red) has a higher concentration of light within $r_{10\%}$ than the metal-poor models.}\label{fig:fig10}
\end{figure}

We observe no size difference with metallicity for the clusters when measuring the size by means of the mass distribution, e.g. half-mass radius. This indicates that the clusters are structurally identical, and different mass-loss rates depending on metallicity are not causing the cluster size to change appreciably. Also, the overall mass and mass segregation are not largely affected by metallicity: a higher MS turnoff mass for the metal-rich cluster is compensated by a lower remnant mass, two effects almost canceling each other out. 
In Fig. \ref{fig:fig9} i) we show the typical radial profile of the average stellar mass 
for the three different metallicities at a late age. 
The models are in good agreement, showing no significant variation with $Z$. 
However we find size differences of up to $20\%$ when measuring the cluster size by means 
of the stellar luminosity. 
The reason for this is two-fold. 
Firstly, less massive remnants in the high-$Z$ cluster free more space in the core for MS and giant 
stars, i.e. luminous matter,  steepening the luminosity profile in the central regions. 
This is evident in Fig. \ref{fig:fig9} ii) which plots the radial profile 
of the average luminosity per radial region. 
The second factor can also be clearly seen in the same figure: 
even though low-$Z$ clusters have a lower MS turnoff mass, the luminosity of MS stars of identical 
masses is higher in the low-$Z$ case. 
This results in the low-$Z$ clusters appearing brighter beyond the centre, with the differences beyond two parsecs being significant in relation to the errorbars, as shown in Fig. \ref{fig:fig9} ii). 
Combined, these effects result in a larger cluster appearance for the metal-poor clusters. 
To reinforce this we show in Fig. \ref{fig:fig10} the luminosity within the $10\%$ Lagrangian radius 
normalized by the total luminosity, as a function of time. 
Here we see that the metal-rich cluster consistently has a greater central concentration 
of luminous matter.

The fact that low-$Z$ stars are brighter for a given mass than their metal-rich counterparts, will be the case independent of a different treatment for NSs and BHs. However, different NS and BH abundances might affect the surface brightness profile by altering the central concentration of luminous stars. We have evolved an additional set of models where NSs and BHs receive a larger kick at formation, resulting in neither sub-population being present in the cluster after a few hundred Myr of cluster evolution (with the exception of the rare case that a NS may form via a WD-WD merger). In contrast to the previous models that contain NSs and BHs, this causes the luminosity profiles for different metallicity clusters to be nearly identical (see the far right panel of Fig. \ref{fig:fig9}). This is no surprise: the remnant mass depends on metallicity and removing the remnants erases some of the metallicity effects. This is in excellent agreement with the findings by \citet{Downing2012}, where significant half-light radii differences are measured with Monte Carlo models (utilizing the same stellar evolution prescription \citealt{SSE}) only when BHs are retained in the cluster. While our model clusters are smaller than those of \citet{Downing2012}, and we only retain a few BHs compared to hundreds in their study, we find the same effect already with very few BHs present, with a contribution also from the NSs that are present. 
 
\subsection{Mass-to-light ratio}\label{ML_ratio}
In Section \ref{SSE_model} we have already mentioned the mass-to-light ratio $M/L$ for a stellar population evolved purely with stellar evolution, but no dynamical interaction (see Fig. \ref{fig:fig2}, right panel). The higher overall luminosity for metal-poor populations implies a lower $M/L$ ratio: the mass-to-light ratio increases with increasing metallicity. The same trend has previously been observed by e.g. \citet{Anders2009b} where \texttt{GALEV} models were computed based on the models of \citet{Baumgardt2003}.
In Fig. \ref{fig:fig4} we repeated the same analysis as in Fig. \ref{fig:fig2}, but now for our $N$-body models. We chose model set b) as an illustrative case, but all three sets are equivalent. The evolution of mass for all metallicities is nearly identical (Fig. \ref{fig:fig4}), whereas the metal-poor cluster has a slightly higher overall mass while the metal rich cluster has a slightly higher MS mass. The overall luminosity is evolving in a similar fashion as in the non-dynamical model, but a factor of two lower owing to the loss of stars. Metallicity differences in $L$ are obvious especially for $t<6\,$Gyr, but continue up to $13$ Gyr.
The dynamical evolution introduces selective effects on the evolution of $M/L$ as low-mass main sequence stars are preferentially lost from the outskirts of the cluster \citep{Baumgardt2003}. Those low-mass stars have a high $M/L$. White dwarfs also have relatively low average mass compared to stars in the central regions. Thus they are candidates to be lost and have a mass-to-light ratio approaching infinity.  As a general rule, losing a low-mass MS star or a white dwarf will \emph{decrease} the mass-to-light ratio (see Table \ref{tb:tb4}). There is an additional effect arising from metallicity differences to consider: for any given mass at a certain time, the luminosity of the metal-poor star will be higher than for a metal-rich star and hence the low-$Z$ star will have a lower $M/L$. This implies that escaping metal-rich stars will cause a larger decrease of $M/L$. In other words: the mass-to-light ratio will be more affected by the loss of low-mass stars in a high-$Z$ cluster. While this is in agreement with the models by \citet{Baumgardt2003} and \citet{Anders2009b}, it is in disagreement with the observed mass-to-light ratios of metal-rich clusters in M$31$ \citep{Strader2009, Strader2011}. \citet{Strader2011} have suggested different initial mass functions for red GCs, which has not been tested here.
\begin{table}
\centering
\caption{Luminosity $L$ and mass-to-light ratio $M/L$ for stars with different masses and metallicities. For given mass, the luminosity increases with metallicity, causing $M/L$ to decrease.}\label{tb:tb4}
\begin{tabular}{ l | c c c}
                    & $0.1\,M_{\odot}$ & $0.5\,M_{\odot}$ &  $0.8\,M_{\odot}$ \\ \hline
$Z=0.01$     & $0.001\,L_{\odot}$  & $0.04\,L_{\odot}$ & $0.32\,L_{\odot}$\\
                    & $100\,M_{\odot}/L_{\odot}$  & $12.5\,M_{\odot}/L_{\odot}$ & $2.5\,M_{\odot}/L_{\odot}$ \\
$Z=0.001$   & $0.0013\,L_{\odot}$ & $0.06\,L_{\odot}$ & $0.5\,L_{\odot}$ \\
                    & $77\,M_{\odot}/L_{\odot}$ & $8.2\,M_{\odot}/L_{\odot}$ & $1.6\,M_{\odot}/L_{\odot}$\\
$Z=0.0001$ & $0.0015\,L_{\odot}$ & $0.07\,L_{\odot}$ & $0.56\,L_{\odot}$ \\
                    &  $66\,M_{\odot}/L_{\odot}$ & $7\,M_{\odot}/L_{\odot}$ & $1.4\,M_{\odot}/L_{\odot}$\\
 \end{tabular}
 \end{table}
 
\section{DISCUSSION AND CONCLUSIONS}
We have measured the sizes of GC models with different metallicity, evolved with the direct $N$-body code \texttt{NBODY6}. All clusters start their evolution with $105\,000$ stars and a mass of $\approx 6 \times 10^4~M_{\odot}$. We find no size differences with metallicity when measuring sizes by means of the half-mass radius or other mass-weighted radii, with the exception that lower remnant masses for high-$Z$ stars cause the $N$-body core radius to fluctuate less. This indicates, that there is no structural difference between clusters of low and high metallicity. Even though the mass-loss rates of low-$Z$ stars are higher, 
especially in the initial stages of evolution, a consequently lower escape velocity and higher average remnant mass cancels this effect, leading to no overall size difference. 
In accordance with this, we also find that the number of stars and cluster mass 
remaining at a particular time do not vary noticeably with the metallicity of the cluster.

\citet{Schulman2012} evolved $N$-body models starting with $N = 8\,192$ stars and different metallicities 
to find a size difference between metal-poor and metal-rich clusters, in terms of the half-mass radius. 
This is in disagreement with our results and those of the Monte Carlo models of \citet{Downing2012}. 
The \citet{Schulman2012} models were evolved with some softening so that the effects of close binaries 
were not included. 
They were evolved to a dynamical age of $5 \, t_{\rm rh}$ which translated to physical ages in 
the range of $100 - 500\,$Myr for the small-$N$ models. 
The claim is that the results should be applicable to larger clusters, including GCs, because the 
impact of different stellar evolution and mass-loss histories at various $Z$ will not depend on $N$, 
and also because they performed models in the range of $1\,024$ to $16\,384$ stars that showed 
similar half-mass radius evolution. 
We would counter that as the MS lifetime of a MS turn-off star changes with age and the half-mass 
relaxation timescale of a cluster varies with $N$, it is not at all obvious that the interplay 
between stellar evolution and cluster dynamics will scale in a straightforward manner.  
Indeed, our models here and the open cluster models of \citet{Hurley2004} with $N \sim 30\,000$, 
both show that the half-mass radius of metal-rich models can be smaller than that of the metal-poor 
models at early times (see Fig. \ref{fig:fig6}) 
but that the difference is erased or even reversed later in the evolution. 
Factors including different core-collapse times, the stellar evolution of low-mass stars as a function 
of metallicity (particularly for globular clusters with ages of $10\,$Gyr or more) 
and different remnant masses need to be taken into account to gain the full picture. 
Furthermore, statistical fluctuations are generally prevalent in small-$N$ simulations and it can be 
necessary to average the results of many instances to establish true behavior 
(e.g. \citealt{Kuepper2008}). 
Our models presented here are at the lower edge of the GC mass function but even for these we 
would suggest that larger models again are desired before making any final judgment about the 
size measurements of GCs in general. 
However, our agreement with the large-scale Monte Carlo models of \citet{Downing2012}, performed 
with $5 \times 10^5$ stars, on the issue of half-mass radius variation (or non-variation) 
with metallicity is reassuring. 

In contrast to the evolution of the half-mass radius, we find that the half-\emph{light} 
(or effective) radius does vary with metallicity. 
We find that blue, metal-poor clusters can appear on average $17\%$ larger than red, metal-rich clusters, 
with even larger differences possible when comparing individual models. 
This is in agreement with observations of extra-galactic GC systems, 
where size differences of $17-30\%$ \citep{Larsen2001, Jordan2005, Woodley2010} have been found. 
It is also in agreement with the Monte Carlo models of \citet{Downing2012}. 
Indeed, our $N$-body models and these Monte Carlo models provide excellent independent validation 
of the main result -- that the observed size differences in GCs are likely caused by the 
interplay of stellar evolution and mass segregation. 
Stellar evolution causes low-$Z$ stars to be brighter than their high-$Z$ counterparts while mass segregation causes the most massive remnants to sink to the centre. Successively more massive remnants in low-$Z$ clusters leads to a steeper surface brightness profile for high-$Z$ clusters. The overall mass segregation is similar for metal-poor and metal-rich clusters but more effective in the luminous stars for high-$Z$ clusters owing to a higher main-sequence turnoff mass. This is in excellent agreement with the predictions of \citet{Jordan2004} using multi-mass Mitchie-King models to estimate the size difference between blue and red GCs, finding a difference of $14\%$ due to the combined effect of mass-segregation and stellar evolution.

The apparent size difference does have a dependence on the treatment of remnants. When ejecting all NSs and BHs, no significant size difference (half-light radius) is found, partly owing to the fact 
that one of the variations with metallicity (remnant masses) has been negated. When we retain $\approx 5\%$ of the NSs and BHs arising from the primordial population, our results are in general agreement with 
the \citet{Downing2012} models that retained large numbers of BHs. While there are uncertainties in the retention fractions for NSs and BHs, there are also uncertainties for the masses of remnant BHs. 
We have used the stellar evolution wing mass-loss prescriptions from \citet{SSE}, while improved, $Z$ dependent mass-loss rates are now available \citep{Vink2001}. However, the resulting differences for BH masses are most apparent for stars above $40\,M_{\odot}$ \citep{Belczynski2010}, while just a few stars are drawn from this mass range in the models presented here.

The average size difference of $17\%$ implies that blue GCs do indeed appear larger as a result 
of metallicity effects. Since this is at the lower end of what is found in observations, 
other causes (such as projection effects) can also be expected to play a role. 
In the future we plan to extend our study by performing additional $N$-body simulations 
that explore parameters such as larger $N$, smaller initial size and differing initial density 
profiles, as well as different cluster orbits, to further understand the effects of cluster 
evolution and environment on measured sizes. 
Our spread of individual measurements in Fig. \ref{fig:fig8} can be compared to extragalactic studies of GC systems as well as in the Milky Way, in which half-light radii of GCs are found to be distributed between $1$ to $8\,$pc (e.g. \citealt{Larsen2003} Fig. $4$, \citealt{Spitler2006} Fig. $19$, \citealt{Madrid2009} Fig. $10$). Since clusters of different masses and at different galactocentric distances are included in the observational samples, a larger scatter is expected than for our models (which currently give values between $2-6\,$pc). We would expect the model spread to increase when we extend our study to include a range of cluster parameters.

In addition to the half-light radius, we have also analyzed the evolution of the mass-to-light ratio. 
When comparing cluster models evolved purely through stellar (but no dynamical) evolution with the thorough $N$-body models, there is little change in $M/L$. As seen before in \citet{Baumgardt2003}, we find that $M/L$ increases with time, where dynamical interactions lead to a decrease in $M/L$ as low-mass stars (carrying a high mass-to-light ratio) are preferentially lost from the cluster. The decrease in overall cluster luminosity with time results in an increase of the mass-to-light ratio.

\section{Acknowledgments}
We thank Marie Martig, Jeremy Webb and Mark Gieles for useful comments and valuable discussion and we thank the anonymous referee for constructive comments. We thank Sverre Aarseth for making \texttt{NBODY6} publicly available. We also thank Dean McLaughlin for providing the \texttt{gridfit} code to us. The simulations were carried out on Tesla S$1070$ graphics cards at Swinburne University. AS thanks the Astronomical Society of Australia for a travel grant that helped to fund a visit to the University of Cambridge and Lorentz Centre in Leiden during this work.

\bibliography{refs}
\end{document}